\documentclass{bmvc2k}

\usepackage{pifont}
\usepackage{booktabs}
\usepackage{multirow}
\usepackage{amsfonts}
\usepackage[table]{xcolor}
\usepackage{float}

\title{SinoDiff: Physics-Consistent Self-Supervised Diffusion for Unified Low-Dose to Standard-Dose PET Sinogram Recovery}

\addauthor{Ghulam Nabi Ahmad Hassan Yar}{ghulam.yar@monash.edu}{1,2}
\addauthor{Himashi Peiris}{himashi.peiris@monash.edu}{1,2}
\addauthor{Sharna Jamadar}{sharna.jamadar@monash.edu}{1,2}
\addauthor{Alex Fornito}{alex.fornito@monash.edu}{1,3}
\addauthor{Zhaolin Chen}{zhaolin.chen@monash.edu}{1,2}

\addinstitution{
 Monash Biomedical Imaging,\\
 Monash University,\\
 Melbourne, Australia
}
\addinstitution{
 Faculty of Information Technology,\\
 Monash University,\\
 Melbourne, Australia
}
\addinstitution{
 Turner Institute for Brain and Mental Health,\\
 School of Psychological Science,\\
 Monash University,\\
 Melbourne, Australia
}

\def\etal{\emph{et al}\bmvaOneDot}
\newcommand{\cmark}{\ding{51}}%
\newcommand{\xmark}{\ding{55}}%
\definecolor{oraclecolor}{HTML}{1BA1E2}
\definecolor{worcolor}{HTML}{2B65B1}
\definecolor{selfcolor}{HTML}{FFF2CC}
\definecolor{proposedcolor}{HTML}{F0A30A}

\newcommand{\oracle}{\rowcolor{oraclecolor!10}}
\newcommand{\wor}{\rowcolor{worcolor!10}}
\newcommand{\self}{\rowcolor{selfcolor!10}}
\newcommand{\proposed}{\rowcolor{proposedcolor!10}}

\runninghead{Ghulam \etal}{SinoDiff}

\begin{document}

\maketitle

\begin{abstract}
Low-dose positron emission tomography (LD-PET) reduces radiation exposure but leads to poor image quality and hinders diagnostic confidence. Existing supervised LD to standard-dose (SD) PET recovery methods often fail to generalise across dose variations, while dose-agnostic supervised methods require paired LD-SD data. Current self-supervised methods, although more flexible, typically produce inferior results, including loss of anatomical details and oversmoothing pathological features. These pose major limitations for practical applications. To overcome these limitations, we propose SinoDiff, a novel self-supervised, physics-consistent diffusion framework for the recovery of PET sinograms across multiple predefined dose levels. Unlike the noise simulation in traditional diffusion methods, SinoDiff integrates the PET acquisition model into the forward diffusion process via Poisson thinning, enabling physically consistent sampling/modelling of dose-dependent count statistics. During the reverse diffusion process, SinoDiff estimates the incremental change in PET signals from predefined dose levels. Therefore, SinoDiff is a single, unified model that requires no retraining across multiple dose levels. To consider the characteristics of PET sinogram, we incorporate a frequency-domain convolution to capture long-range dependencies across projection angles and detector bins. Experiments on $^{18}$F-FDG and $^{18}$F-FDOPA datasets demonstrate that SinoDiff achieves competitive performance against supervised and self-supervised baselines across multiple dose levels. The code is available at: \href{https://github.com/Gnahy/SinoDiff.git}{SinoDiff.}
\end{abstract}

\section{Introduction}
\label{sec:intro}

Positron Emission Tomography (PET) is critically important for the diagnosis and prognosis of oncological, neurological, and cardiological conditions. The PET imaging process begins with the administration of a radioactive tracer, which distributes throughout the body according to the tracer's biochemical properties and the metabolic activity of different tissues. Over time, the radionuclide undergoes positron decay, producing positrons that annihilate with nearby electrons and generate pairs of 511~keV photons. These coincident photons are detected by the PET scanner and recorded as spatiotemporal list-mode data~\cite{pain2022deep}. The list-mode data are then histogrammed into sinograms, which are subsequently reconstructed into PET images using algorithms such as ordered subsets expectation maximisation (OSEM)~\cite{hudson1994accelerated}. However, PET relies on radioactive tracers and radiation exposure, which raises concerns, particularly for paediatric patients and individuals requiring repeated scans~\cite{kunnapuu2025radiation}. Acquiring PET at a reduced dose level mitigates the risk. However, low-dose (LD) PET suffers from signal loss, resulting in poor image quality. This reduced dose may correspond to low-dose or ultra-low-dose PET acquisitions, where only a fraction of the standard-dose (SD) counts are available, such as 20\%, 10\%, or even 5\% of the original acquisition counts. Thus, recovering SD PET from LD PET is a key research question, aiming to recover signal fidelity and image quality while maintaining the benefits of reduced radiation exposure~\cite{10.1007/978-3-031-72104-5_36}.

\begin{figure}
    \centering
    \includegraphics[width=\textwidth]{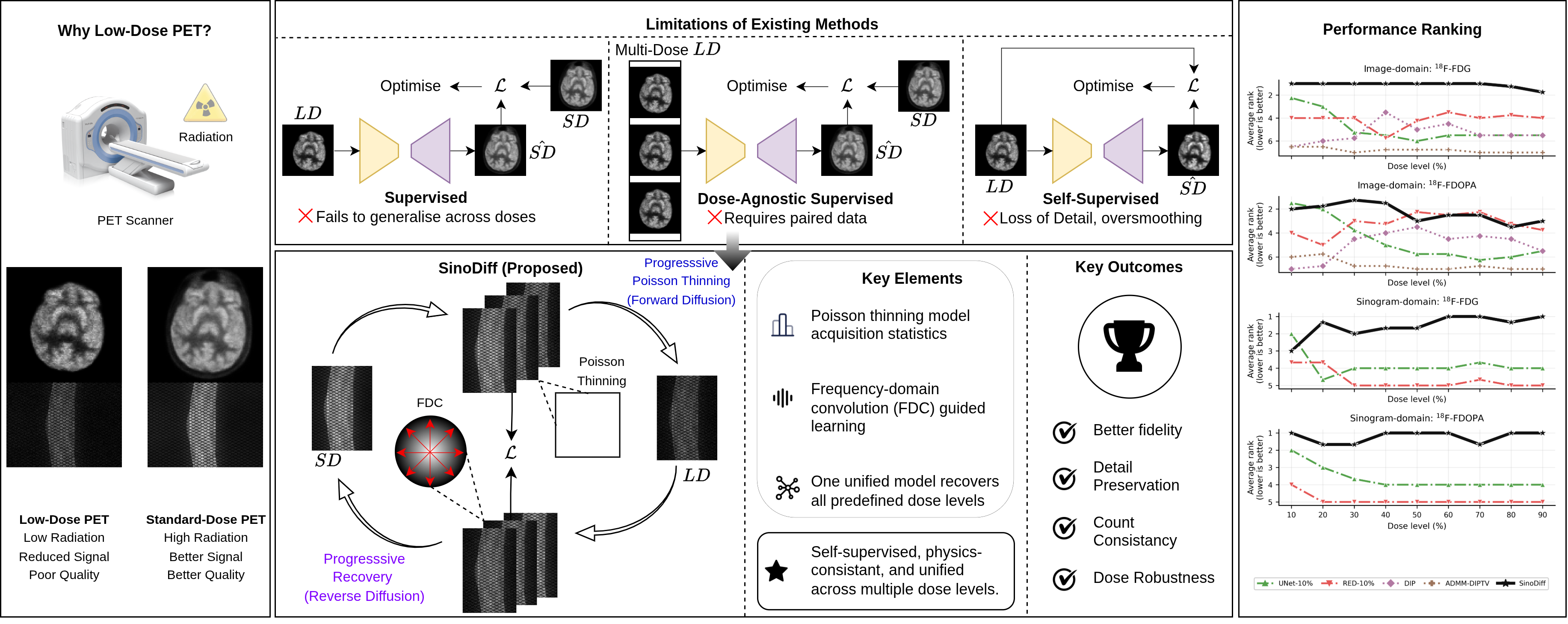}
    \caption{Low-dose PET reduces radiation exposure but degrades image quality. The proposed SinoDiff model uses self-supervised sinogram-domain diffusion with Poisson thinning and frequency-domain convolution to recover standard-dose PET without paired LD–SD data, improving dose generalisation.}
    \label{fig:teaser}
\end{figure}

Various image-based and sinogram-based supervised deep learning methods have been developed for LD-to-SD PET recovery~\cite{chen2020generalization, zeng2025mak, 10.1007/978-3-031-72104-5_52, pain2024deep, yar2026m2diff}. However, these methods are dose-specific, demonstrate poor generalisability, incur increased computational cost, and have limited overall clinical utility. 
To improve generalisability, dose-aware LD-to-SD PET recovery has been proposed~\cite{cui2024s3pet, fei2024two, sudarshan2021towards}. However, these methods rely on paired supervision and multimodal inputs such as MRI. 
To alleviate the dependence on paired training data, self-supervised image domain approaches such as Noise2Noise~\cite{kang2021noise2noise, chan2019noise}, Noise2Void~\cite{song2021noise2void}, and Deep Image Prior (DIP)~\cite{gong2018pet, onishi2023self} have been proposed. These methods avoid explicit labelled supervision by deriving training signals directly from the data itself, typically by predicting one part or view of the input from another~\cite{rani2023self, jiang2023pet}. Although these methods are effective for denoising, their performance is often inferior to fully supervised approaches, and they have a tendency to overfit to noise~\cite{nittscher2024svd, shi2022measuring}. Inconsistent performance becomes more pronounced under real-world clinical variability~\cite{chaudhari2021low}. This issue is largely due to inadequate modelling of real PET signal statistics, such as Poisson counting statistics and physics-consistent dose modelling. 

To address the above limitations, we propose a self-supervised diffusion framework, termed SinoDiff, for PET sinogram recovery that incorporates PET acquisition physics and is trained only on SD PET data, as illustrated in Fig.~\ref{fig:teaser}. SinoDiff is generalisable to predefined dose levels without retraining the model. Our main contributions include: 
\begin{enumerate}
    \item A novel physics-consistent self-supervised diffusion framework that integrates the PET Poisson data model into the forward diffusion process, which ensures strong alignment between the diffusion sampling trajectory and the real data statistical process.
    \item A unified inference model allowing a single trained model to generalise across multiple predefined dose levels. 
    \item A Frequency Domain Convolution (FDC) kernel to model global projection correlations in the sinogram domain. By operating in the Fourier domain, FDC captures long-range dependencies across projection angles and detector bins that are disrupted by Poisson thinning and are difficult to recover using local spatial convolutions. 
\end{enumerate}
\section{Related Work}
\label{sec:related}

Recent deep learning approaches for LD PET recovery can be categorised into supervised image-domain methods, self-supervised methods, and sinogram-domain and physics-aware recovery techniques.

\paragraph{Supervised Methods}
Supervised image-domain methods have been widely explored for LD-to-SD PET recovery using UNets~\cite{chen2020generalization}, GANs~\cite{zhao2020study}, Transformers~\cite{kaviani2023image}, and diffusion models~\cite{huang2025diffusion,10.1007/978-3-031-72104-5_52}. More recent dose-aware frameworks incorporate dose conditioning to improve robustness across varying count levels~\cite{Xie2020dose,tang2026halo,azimi2025deep}. Although these approaches achieve strong denoising performance, they generally rely on paired LD-SD supervision and often require retraining or explicit conditioning for different dose settings. In contrast, our proposed method does not require paired data, as it generates its own training signals.

\paragraph{Self-Supervised Methods}
To reduce dependence on paired supervision, self-supervised methods such as Noise2Noise~\cite{kang2021noise2noise,chan2019noise}, Noise2Void~\cite{song2021noise2void}, and Deep Image Prior (DIP)~\cite{gong2018pet,onishi2023self} have been explored for PET denoising. Jiang \etal~\cite{10.1007/978-3-031-43907-0_1} proposed a self-supervised latent diffusion model that only takes SD PET images. They adapted the model using Poisson noise modelling for LD PET recovery. Other approaches often incorporate anatomical priors or multimodal guidance, including MRI~\cite{onishi2023self,cui2021populational} and attenuation maps~\cite{li2025fastdip}, to stabilise optimisation and improve restoration quality. However, recent self-supervised methods rely heavily on the DIP backbone, often exhibit unstable optimisation, rely on auxiliary priors, or do not explicitly model PET acquisition statistics within the diffusion process. On the contrary, our proposed model leverages the physics of PET acquisition to generate strong training signals that stabilise the output without requiring anatomical priors.

\paragraph{Physics-aware and Sinogram Domain Methods}
Several studies have explored PET recovery directly in the sinogram domain to better preserve acquisition-space information. Leffler \etal~\cite{leffler2026filling} proposed a residual U-Net operating on sinograms, while Ai \etal~\cite{ai2025red} replaced Gaussian diffusion noise with residual-based sinogram modelling. Physics-aware PET recovery methods further incorporate uncertainty estimation~\cite{sudarshan2021towards, cui2022pet} or noise modeling~\cite{li2022noise, sanaei2023employing, xie2023unified} to improve PET fidelity. However, existing methods remain predominantly supervised, and limited attention has been given to embedding PET acquisition statistics directly within the recovery process. Unlike approaches that incorporate PET physics primarily through sinogram inputs, conditioning, or external priors, SinoDiff addresses this gap by embedding Poisson thinning directly into a self-supervised, physics-consistent sinogram diffusion process.
\section{Methodology}
\label{sec:method}

\subsection{Problem Formulation}

Throughout the paper, vectors are denoted by bold lower-case letters $\mathbf{x}$, while matrices and tensors are denoted by bold upper-case letters $\mathbf{X}$. Let $\mathcal{X}$ denote the set of SD PET sinograms, where each sample $\mathbf{X} \in \mathbb{R}^{C \times H \times W}$ 
represents a multi-slice input.

\paragraph{PET Data Model.}
Let $\boldsymbol{\Lambda}(\mathbf{\lambda}) \in \mathbb{R}_+^{M}$ denote the expected sinogram rate induced by PET tracer activity $\mathbf{\lambda}$. PET acquisition over duration $\tau$ can be modelled as a Poisson counting process, yielding the list-mode/sinogram data, $\mathbf{X}$.
\begin{equation}
\boldsymbol{\Lambda}(\mathbf{\lambda}) 
= \mathbf{N}\!\left(\mathbf{A}\mathbf{\lambda}\right) + \mathbf{b}, \quad \mathbf{X} \sim \mathrm{Poisson}(\tau\,\boldsymbol{\Lambda}(\mathbf{\lambda})).
\label{eq:pet_poisson}
\end{equation}
where $\mathbf{A}$ is the PET system projection operator, $\mathbf{N}(\cdot)$ accounts for attenuation and normalisation effects, and $\mathbf{b}$ denotes the combined background term including scatter and random events.

\paragraph{Low-Dose PET Recovery.}
The primary objective is to learn a mapping function $\mathcal{G}$ that recovers the SD PET from the LD measurements in the sinogram domain.  Given an LD PET sinogram, $\mathbf{X}_{LD}$, this mapping is defined as:
\begin{equation}
    \mathbf{X}_{SD} = \mathcal{G}(\boldsymbol{\Theta}; \mathbf{X}_{LD}),
\end{equation}
where $\mathcal{G}(\cdot)$ denotes the estimated recovery network, and $\boldsymbol{\Theta}$ represents its learnable parameter set. The $\boldsymbol{\Theta}$ combines a data-driven term and a physics-consistent term penalising disagreement between the estimated and the acquired PET signals~\cite{pain2022deep}. Unlike supervised LD-SD recovery methods, SinoDiff does not rely on paired independently acquired LD and SD scans. Instead, it constructs self-supervised training pairs by progressively degrading SD PET sinograms through a physics-consistent Poisson thinning process, producing nested low-count observations at predefined dose levels. By modelling the underlying Poisson counting statistics directly in the sinogram domain, SinoDiff implicitly enforces PET count consistency and enables dose-generalised recovery without dose-specific retraining.


\begin{figure}
    \centering
    \includegraphics[width=\linewidth]{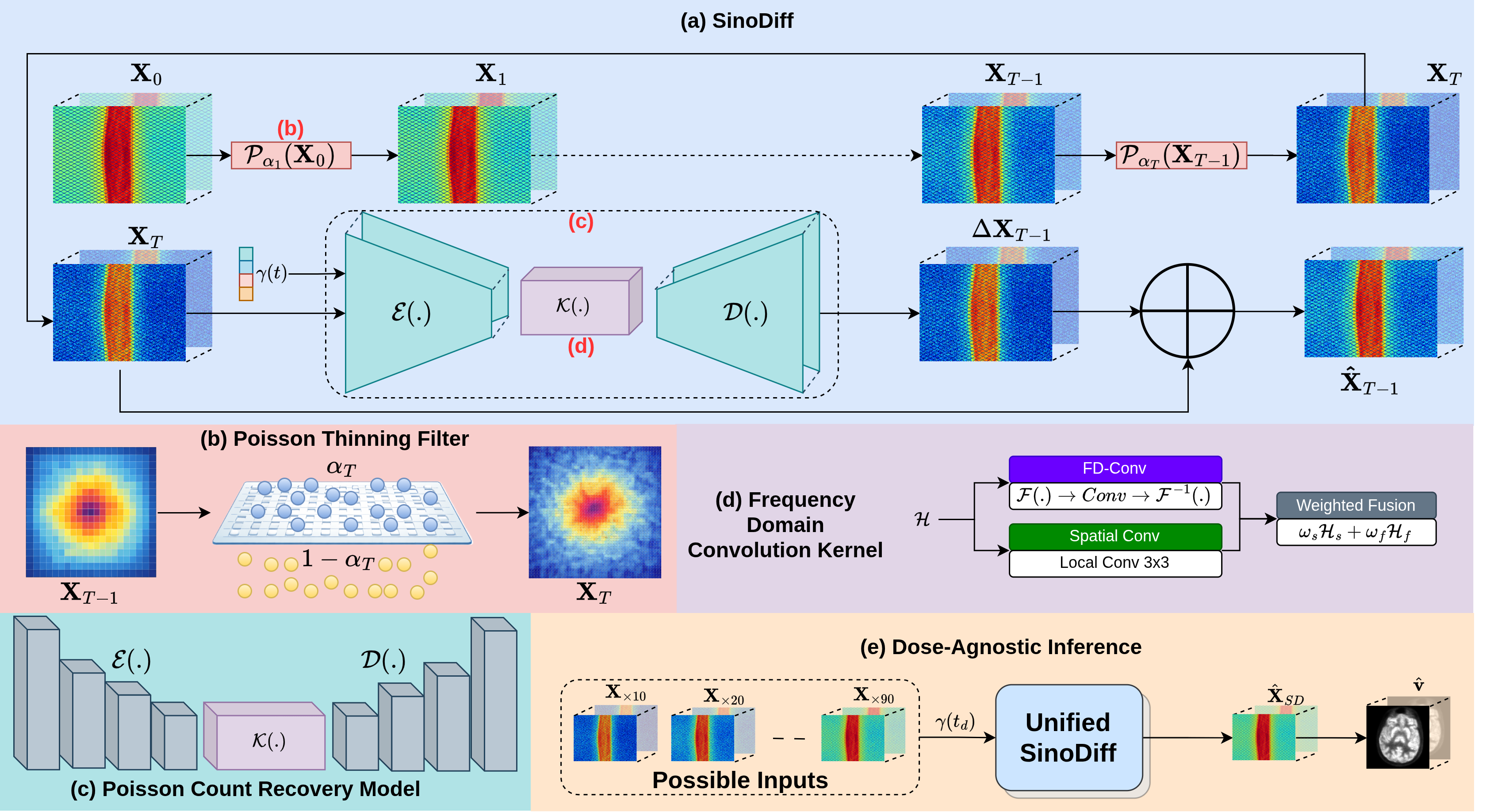}
    \caption{Overall workflow of SinoDiff. (a) forward-reverse dose recovery, (b) Poisson thinning-based count reduction, (c) encoder-FDC-decoder recovery network, (d) frequency-domain modelling of global sinogram dependencies, and (e) unified inference across dose levels.} 
    \label{fig:architecture}
\end{figure}

\subsection{SinoDiff Model}
The proposed SinoDiff framework is built on the principles of diffusion modelling and introduces PET-specific components that align the degradation and recovery processes with sinogram-space count statistics. As illustrated in Fig.~\ref{fig:architecture}, the proposed framework consists of five main building blocks: (a) a complete forward and reverse dose-recovery workflow, (b) a physics-consistent Poisson thinning operator for simulating progressive count reduction, (c) a progressive recovery network composed of an encoder $\mathcal{E}(\cdot)$, FDC kernel $\mathcal{K}(\cdot)$, and decoder $\mathcal{D}(\cdot)$, (d) a frequency-domain convolution module for modelling global sinogram dependencies, and (e) a unified inference strategy that uses a single trained model across different dose levels.

\paragraph{Forward Diffusion Process.}
Unlike conventional Gaussian diffusion, SinoDiff's forward process follows Poisson counting statistics governed by the PET physics. In SinoDiff, dose degradation is modelled as a discrete Markov chain in the sinogram domain:
\begin{equation}
\mathbf{X}_0 \rightarrow \mathbf{X}_1 \rightarrow \cdots \rightarrow \mathbf{X}_T,
\end{equation}
where $t \in \{1,\dots,T\}$ is the diffusion timestep. In SinoDiff, $t$ reflects tracer dose levels
and since each timestep corresponds to a progressively lower count level, we refer to $t$ as the \emph{dose-step} throughout this work.
The forward process is factorised as:
\begin{equation}
q(\mathbf{X}_{1:T}\mid \mathbf{X}_0)
= \prod_{t=1}^{T} q(\mathbf{X}_t \mid \mathbf{X}_{t-1}),
\end{equation}
with each transition representing an independent thinning of data statistics. At dose-step $t$, each detected data/event in $\mathbf{X}_{t-1}$ is retained with probability $\alpha_t \in (0,1)$:
\begin{equation}
\mathbf{X}_{t}
=
\mathcal{P}_{\alpha_t}(\mathbf{X}_{t-1}),
\label{eq:binomial_thin}
\end{equation}
where $\mathcal{P}_{\alpha_t}(\cdot)$ denotes the Poisson thinning operator.

Let $\bar{\alpha}_t = \prod_{k=1}^{t}\alpha_k$ denote the cumulative retention probability of detected events. By the Poisson thinning property and the PET acquisition model in Eq.~\eqref{eq:pet_poisson}, the marginal distribution remains Poisson:
\begin{equation}
\mathbf{X}_t \sim \mathrm{Poisson}\!\left(\bar{\alpha}_t\,\tau\,\boldsymbol{\Lambda}(\mathbf{\lambda})\right).
\label{eq:poisson_scaled}
\end{equation}
The above formulation is consistent with the PET data acquisition process and eliminates potential mismatches between the actual acquired PET data and sampled diffusion signals. The predefined thinning schedule provides a discretised approximation of progressive count reduction, in which each diffusion step corresponds to a controlled reduction in detected events via Poisson thinning. Since PET dose reduction primarily manifests as reduced count statistics, the thinning process preserves the acquisition-consistent behaviour of low-dose PET measurements.

\paragraph{Progressive PET Dose Recovery.} Given $\mathbf{X}_t \sim \mathrm{Poisson}(\bar{\alpha}_t\,\tau\,\boldsymbol{\Lambda}(\mathbf{\lambda}))$, the reverse dose recovery step aims to recover a higher-dose signal
\begin{equation}
\mathbf{X}_{t-1} \sim \mathrm{Poisson}(\bar{\alpha}_{t-1}\,\tau\,\boldsymbol{\Lambda}(\mathbf{\lambda})),
\quad \bar{\alpha}_{t-1} > \bar{\alpha}_t.
\label{eq:reverse_target_dist}
\end{equation}
Due to the additivity of Poisson processes, the higher-dose sinogram can be decomposed as:
\begin{equation}
\mathbf{X}_{t-1} = \mathbf{X}_t + \Delta \mathbf{X}_{t-1},
\label{eq:dose_estimation}
\end{equation}
where $\Delta \mathbf{X}_{t-1}$ denotes the increment of counts removed during thinning. Consequently, reverse dose recovery reduces to estimating the missing Poisson increment $\Delta \mathbf{X}_{t-1}$ conditioned on the observed $\mathbf{X}_t$.

Since the underlying tracer activity $\lambda$ is unknown, we approximate the conditional expectation $\mathbb{E}[\Delta \mathbf{X}_{t-1} \mid \mathbf{X}_t]$ using a learnable model $\mathcal{G}(\boldsymbol{\Theta}; \mathbf{X}_t, \gamma(t))$, where $\gamma(t)$ represents sinusoidal embedding of $t$. The predicted residual $\Delta \mathbf{X}_{t-1}$ thus estimates the expected missing counts:
\begin{equation}
\Delta\mathbf{X}_{t-1}=\mathcal{G}(\boldsymbol{\Theta}; \mathbf{X}_{t}, \gamma(t)); 
\label{eq:gen_model}
\end{equation}
This process is iteratively applied for all $t$ to obtain the final estimate $\hat{\mathbf{X}}_{0}$.
To design $\mathcal{G}$ operating in the sinogram domain, we consider the intrinsic difference of a projection/sinogram domain compared to the conventional image domain. In the sinogram domain, anatomical structures induce correlated patterns across many projection angles that are disrupted by thinning, rendering local-only convolutions ineffective, especially at very low dose levels. 
To address this, inspired by~\cite{banerjee2024fcdm}, we introduce an FDC (Frequency Domain Convolution) kernel that performs learnable Fourier-domain filtering to capture long-range correlations across projection bins, thereby enabling restoration of globally consistent sinogram structure. Given $\mathbf{\mathcal{H}}$ as output of $\mathbf{\mathcal{E}(.)}$, the FDC kernel can be written as:
\begin{equation}
\mathbf{\mathcal{K}(\mathcal{H})} =
\omega_s \mathbf{\mathcal{H}_{s}} +
\omega_f \, \mathcal{F}^{-1}\!\left(
\mathbf{W}_f \odot \mathcal{F}(\mathbf{\mathcal{H}})
\right),
\end{equation}
where $\mathcal{F}(\cdot)$ and $\mathcal{F}^{-1}(\cdot)$ denote the forward and inverse Fourier transforms respectively, $\mathbf{W}_f$ is a learnable complex mixing matrix, $\odot$ represents element-wise multiplication, $\mathbf{\mathcal{H}_s}$ represents the output of the spatial convolution branch, and $\omega_s,\omega_f$ are learnable fusion weights for the spatial and frequency branches, respectively.

\paragraph{Objective Function.} Although PET acquisition follows discrete Poisson statistics, SinoDiff estimates the continuous expectation of sinogram intensities rather than discrete event realisations. Therefore, continuous regression losses, such as MSE, are used for expected-count recovery, while Poisson thinning preserves acquisition-consistent degradation during forward diffusion.

The model is trained in a self-supervised manner by comparing the $\hat{\mathbf{X}}_{t-1}$ with the target $\mathbf{X}_{t-1}$ acquired during the physics-consistent forward process. The total loss is composed of three complementary terms
\begin{equation}
\begin{split}
\mathcal{L} =
\mathcal{L}_{MSE}
+ \beta_{\mathrm{SSIM}} \mathcal{L}_{SSIM}
+ \beta_{\mathrm{Freq}} \mathcal{L}_{Freq}.
\end{split}
\label{eq:total_loss}
\end{equation}
Here, $\mathcal{L}_{MSE}$ is the mean squared error loss, $\mathcal{L}_{SSIM}$ is the structural similarity loss, and $\mathcal{L}_{Freq}$ measures the $L_2$ distance in the Fourier domain, with $\beta_{\mathrm{SSIM}}$ and $\beta_{\mathrm{Freq}}$ controlling their weights.

\paragraph{Unified Inference.} During inference, a predefined low-dose sinogram $\mathbf{X}_{\mathrm{LD}}$ and its corresponding dose-step $t_{\mathrm{d}}$ are provided as input. The trained model then performs iterative recovery as follows,
\begin{equation}
\hat{\mathbf{X}}_{\mathrm{SD}} = \mathcal{G}(\boldsymbol{\Theta}_{t_d}; \mathbf{X}_{{t_{d}}}, \gamma(t_{d})),
\end{equation}
where $\boldsymbol{\Theta}_{t_d}$ denotes the subset of parameters involved in the reverse steps from $t_d$ to $1$.
Finally, PET scans are reconstructed from the recovered SD PET sinogram using OSEM.

\subsection{Sinogram Domain Count Consistency Analysis}

To assess whether the recovered sinograms preserve PET count statistics, we evaluate count consistency at global, angular, and radial levels. Let $\mathbf{X}_{\mathrm{SD}} \in \mathbb{R}^{C \times H \times W}$ denote the standard-dose reference sinogram and $\hat{\mathbf{X}}_{\mathrm{SD}} \in \mathbb{R}^{C \times H \times W}$ denote the recovered sinogram, where $C$, $H$, and $W$ correspond to slices, projection angles, and detector/radial bins, respectively.

Global count bias (GCB) measures total count preservation:
\begin{equation}
\mathrm{GCB}(\%) =
100
\frac{
\sum_{c,h,w}\hat{X}_{\mathrm{SD},c,h,w}
-
\sum_{c,h,w}X_{\mathrm{SD},c,h,w}
}{
\sum_{c,h,w}X_{\mathrm{SD},c,h,w}
}.
\label{eq:gcb}
\end{equation}
Values closer to zero indicate better total count preservation.

Angular count bias (ACB) evaluates count consistency across projection angles:
\begin{equation}
\mathrm{ACB}_{h}(\%) =
100
\frac{
\sum_{c,w}\hat{X}_{\mathrm{SD},c,h,w}
-
\sum_{c,w}X_{\mathrm{SD},c,h,w}
}{
\sum_{c,w}X_{\mathrm{SD},c,h,w}
},
\quad
\mathrm{Mean}|\mathrm{ACB}| =
\frac{1}{H}\sum_{h=1}^{H}|\mathrm{ACB}_{h}|.
\label{eq:acb}
\end{equation}

Similarly, radial count bias (RCB) evaluates count consistency across detector/radial bins:
\begin{equation}
\mathrm{RCB}_{w}(\%) =
100
\frac{
\sum_{c,h}\hat{X}_{\mathrm{SD},c,h,w}
-
\sum_{c,h}X_{\mathrm{SD},c,h,w}
}{
\sum_{c,h}X_{\mathrm{SD},c,h,w}
},
\quad
\mathrm{Mean}|\mathrm{RCB}| =
\frac{1}{W}\sum_{w=1}^{W}|\mathrm{RCB}_{w}|.
\label{eq:rcb}
\end{equation}
Lower Mean $|\mathrm{ACB}|$ and Mean $|\mathrm{RCB}|$ indicate more consistent count preservation across projection angles and detector bins, respectively.
\section{Experiments \& Results}
\label{sec:experiments}

\paragraph{\textbf{Datasets.}}
Experiments were conducted on two clinical PET datasets acquired on a 3T Siemens Biograph mMR PET/MR scanner.
The \textbf{$^{\textbf{18}}$F-FDG dataset}~\cite{10.1093/gigascience/giac031} includes 27 subjects (250~MBq, 60-min acquisition), while the \textbf{$^{\textbf{18}}$F-FDOPA dataset} comprises 25 subjects (152.17~MBq, 60-min acquisition). For both datasets, raw list-mode data were converted to sinograms without axial compression, scatter, or attenuation correction to preserve native count statistics, yielding 837 slices per subject (43,524 total). These SD sinograms are scanner-acquired rather than synthetic, while LD sinograms are generated through PET Poisson thinning during SinoDiff's forward process. The sinograms were divided into overlapping 2.5D samples using a window size of 9, a stride of 7, and an edge size of 1, yielding approximately 120 samples per subject. The data were split at the subject level. For training, the $^{18}$F-FDOPA and $^{18}$F-FDG datasets used 19 subjects (2,280 samples) and 20 subjects (2,400 samples), respectively. Both datasets used 4 subjects for testing (480 samples each). For validation, $^{18}$F-FDOPA used 2 subjects (240 samples), while $^{18}$F-FDG used 3 subjects (360 samples).
For reconstruction of LD, SD, and recovered sinograms, PET images were reconstructed using NiftyPET with MR-derived $\mu$-maps for attenuation and scatter correction, employing OSEM.

\paragraph{\textbf{Implementation Details.}}
SinoDiff was implemented as a 2.5D model in PyTorch and trained end-to-end on an NVIDIA A40 GPU for 100 epochs using Adam optimiser (LR = $10^{-4}$). The diffusion length was set to $T=10$, corresponding to 10\% dose reduction per step, with loss weights $\beta_{\mathrm{SSIM}}=\beta_{\mathrm{Freq}}=0.2$.

\begin{figure}
    \centering
    \includegraphics[width=\linewidth]{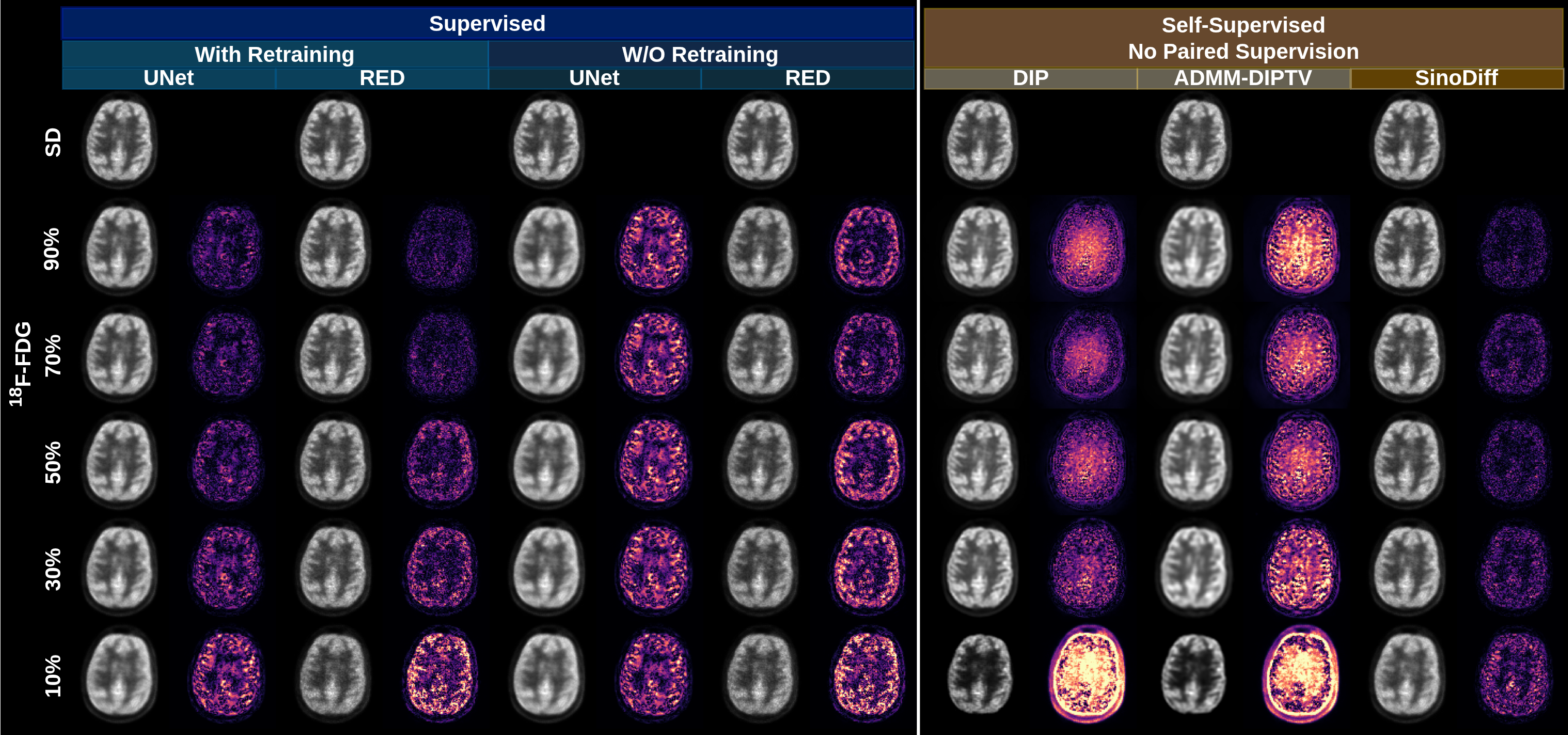}
    \caption{Qualitative comparison of supervised methods with and without retraining, self-supervised methods, and proposed SinoDiff on  $^{18}$F-FDG across dose levels. Grayscale: reconstructions (range: 0.0-1.0), and colour maps: absolute errors (range: 0.0–0.18).}
    \label{fig:dose_comp_dacra_fdg}
\end{figure}

\begin{figure}
    \centering
    \includegraphics[width=\linewidth]{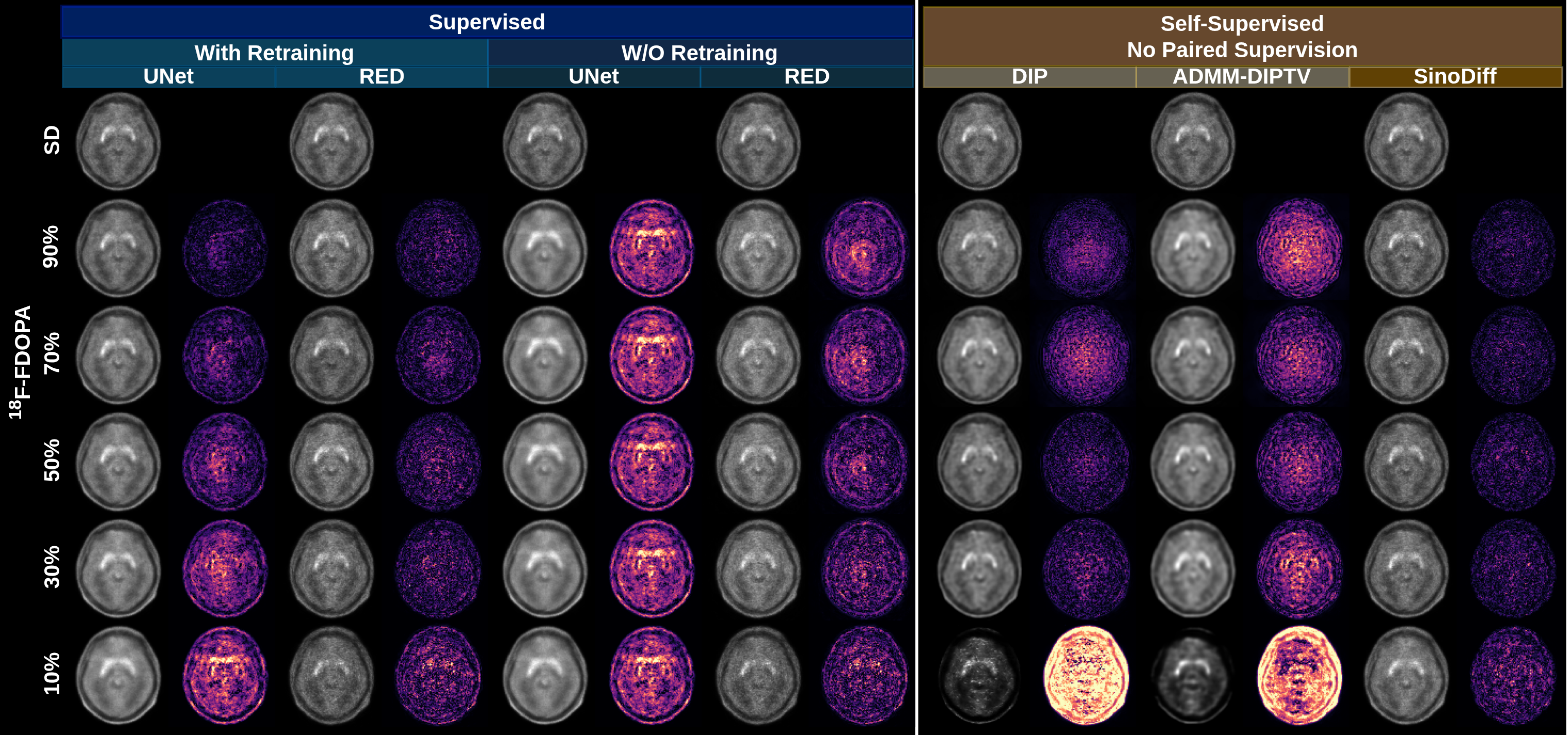}
    \caption{Qualitative comparison of supervised methods with and without retraining, self-supervised methods, and proposed SinoDiff on $^{18}$F-FDOPA across dose levels. Grayscale: reconstructions (range: 0.0–1.0); colour maps: absolute error (range: 0.0–0.22).}
    \label{fig:dose_comp_dacra_fdopa}
\end{figure}

\paragraph{\textbf{Comparative Experiments.}}
To evaluate SinoDiff, we compared it against four baselines covering supervised and self-supervised PET recovery paradigms. The selected methods represent widely used PET denoising frameworks and recent sinogram-domain diffusion approaches. Self-supervised image-domain baselines include: 1) Deep Image Prior (DIP)~\cite{gong2018pet}, a commonly used self-supervised PET denoising framework, and 2) ADMM-DIPTV~\cite{cascarano2021combining}, which extends DIP with total variation regularisation via Alternating Direction Method of Multipliers (ADMM). Supervised sinogram-domain baselines include: 1) Residual UNet~\cite{leffler2026filling}, a supervised sinogram restoration framework, and 2) Residual Estimation Diffusion (RED)~\cite{ai2025red}, a recent diffusion-based sinogram denoising method. These sinogram-domain methods provide the closest controlled comparison to SinoDiff, while their paired LD–SD training provides stronger supervision than the SD-only self-supervised setting used by SinoDiff.

Supervised models were evaluated under two setups: \textbf{(1)} retrained and evaluated at each dose level, \textbf{(2)} without retraining, trained at 10\% dose level and tested across all other predefined doses. Self-supervised methods, DIP and ADMM-DIPTV, learn directly from the LD PET volume without separate training.
In contrast, SinoDiff follows an SD-only self-supervised setting and was trained once on SD sinograms and tested on all predefined dose levels, demonstrating its generalisability. After sinogram enhancement, PET images were reconstructed for evaluation using 
Structural Similarity Index Measure (SSIM), Peak Signal-to-Noise Ratio (PSNR), Normalised Mean Squared Error (NMSE), and Learned Perceptual Image Patch Similarity (LPIPS).

\paragraph{Image Domain Analysis:} The qualitative results in Figs.~\ref{fig:dose_comp_dacra_fdg}, and~\ref{fig:dose_comp_dacra_fdopa} show that supervised baselines with retraining produce sharper images and decrease errors compared with those without retraining. This demonstrates the limited generalisability of supervised methods. The self-supervised baselines, including DIP and ADMM-DIPTV, at a lower dose (10\%) demonstrate severe structural loss and remain inconsistent across dose levels despite partial improvement at higher counts. This behaviour arises because the performance of self-supervised methods relies on early stopping and hyperparameter tuning. Compared with both supervised baselines, SinoDiff shows competitive performance, and when compared to self-supervised baselines, SinoDiff shows superior anatomical details and produces the lowest errors across all predefined dose levels without retraining.



Table~\ref{tab:fdg_all_dose} and~\ref{tab:fdopa_all_dose} summarise quantitative results for $^{18}$F-FDG and $^{18}$F-FDOPA across all predefined dose levels respectively. Supervised models retrained independently for each dose level generally achieve strong performance, particularly for the $^{18}$F-FDOPA dataset. However, their performance degrades noticeably when evaluated in the cross-dose setting without retraining. In contrast, SinoDiff maintains stable performance across all predefined dose levels using a single unified model and consistently achieves the best or second-best results among all methods. For the $^{18}$F-FDG dataset, SinoDiff achieves the highest SSIM and PSNR and the lowest NMSE and LPIPS across all dose levels. For the $^{18}$F-FDOPA dataset, SinoDiff remains competitive with supervised models and outperforms self-supervised baselines. In comparison, image-domain self-supervised methods exhibit unstable optimisation and sharp performance degradation at ultra-low-dose levels. Paired two-tailed t-tests across subject-dose pairs showed that SinoDiff significantly outperformed all UNet and RED settings across all metrics on FDG (p<0.05). On FDOPA, SinoDiff significantly outperformed RED with retraining across all metrics and RED without retraining for SSIM and LPIPS; differences in PSNR and NMSE were not significant. Against UNet with retraining, only SSIM differed significantly. These results highlight the advantage of integrating physics-consistent diffusion and sinogram-domain modelling for cross-dose PET recovery.

\begin{table}[t]
\centering
\caption{
Comparison of image-domain metrics between supervised methods with and without retraining,
self-supervised methods, and proposed SinoDiff across multiple dose levels for the
$^{18}$F-FDG dataset.
}
\label{tab:fdg_all_dose}
\setlength{\tabcolsep}{2.5pt}

\resizebox{\textwidth}{!}{
\begin{tabular}{llllccccccccc}
\toprule
& & &
& \multicolumn{9}{c}{$^{18}$F-FDG Dose (\%)} \\
\cmidrule(lr){5-13}
Metric & Training setting & Backbone & Method
& 90 & 80 & 70 & 60 & 50 & 40 & 30 & 20 & 10 \\
\midrule

\oracle
\cellcolor{white}
& With Retraining & UNet & UNet
& 0.9610 $\pm$ \scriptsize{0.0037}
& 0.9683 $\pm$ \scriptsize{0.0023}
& 0.9631 $\pm$ \scriptsize{0.0053}
& 0.9601 $\pm$ \scriptsize{0.0045}
& 0.9591 $\pm$ \scriptsize{0.0060}
& 0.9482 $\pm$ \scriptsize{0.0056}
& 0.9477 $\pm$ \scriptsize{0.0048}
& 0.9291 $\pm$ \scriptsize{0.0073}
& 0.9156 $\pm$ \scriptsize{0.0060} \\

\oracle
\cellcolor{white}
& With Retraining & Diffusion & RED
& 0.9714 $\pm$ \scriptsize{0.0038}
& 0.9631 $\pm$ \scriptsize{0.0113}
& 0.9660 $\pm$ \scriptsize{0.0076}
& 0.9413 $\pm$ \scriptsize{0.0455}
& 0.9553 $\pm$ \scriptsize{0.0145}
& 0.9415 $\pm$ \scriptsize{0.0060}
& 0.9407 $\pm$ \scriptsize{0.0112}
& 0.8369 $\pm$ \scriptsize{0.1245}
& 0.7505 $\pm$ \scriptsize{0.0945} \\

\cmidrule(lr){2-13}

\wor
\cellcolor{white}
& W/O Retraining & UNet & UNet
& 0.9241 $\pm$ \scriptsize{0.0042}
& 0.9237 $\pm$ \scriptsize{0.0054}
& 0.9237 $\pm$ \scriptsize{0.0046}
& 0.9236 $\pm$ \scriptsize{0.0036}
& 0.9226 $\pm$ \scriptsize{0.0048}
& 0.9222 $\pm$ \scriptsize{0.0057}
& 0.9214 $\pm$ \scriptsize{0.0046}
& \underline{0.9187 $\pm$ \scriptsize{0.0067}}
& \underline{0.9156 $\pm$ \scriptsize{0.0060}} \\

\wor
\cellcolor{white}
& W/O Retraining & Diffusion & RED
& \underline{0.9578 $\pm$ \scriptsize{0.0043}}
& \underline{0.9538 $\pm$ \scriptsize{0.0058}}
& \underline{0.9525 $\pm$ \scriptsize{0.0060}}
& \underline{0.9430 $\pm$ \scriptsize{0.0118}}
& \underline{0.9439 $\pm$ \scriptsize{0.0068}}
& 0.9253 $\pm$ \scriptsize{0.0178}
& \underline{0.9289 $\pm$ \scriptsize{0.0099}}
& 0.8657 $\pm$ \scriptsize{0.0566}
& 0.7505 $\pm$ \scriptsize{0.0945} \\

\self
\cellcolor{white}
& Self-supervised & DIP & DIP
& 0.9407 $\pm$ \scriptsize{0.0105}
& 0.9429 $\pm$ \scriptsize{0.0090}
& 0.9506 $\pm$ \scriptsize{0.0076}
& 0.9496 $\pm$ \scriptsize{0.0098}
& \underline{0.9510 $\pm$ \scriptsize{0.0117}}
& \underline{0.9467 $\pm$ \scriptsize{0.0177}}
& 0.8988 $\pm$ \scriptsize{0.0409}
& 0.6806 $\pm$ \scriptsize{0.0937}
& 0.3384 $\pm$ \scriptsize{0.0522} \\

\self
\cellcolor{white}
& Self-supervised & DIP & ADMM-DIPTV
& 0.9131 $\pm$ \scriptsize{0.0169}
& 0.9196 $\pm$ \scriptsize{0.0105}
& 0.9211 $\pm$ \scriptsize{0.0135}
& 0.9215 $\pm$ \scriptsize{0.0119}
& 0.9228 $\pm$ \scriptsize{0.0208}
& 0.9218 $\pm$ \scriptsize{0.0218}
& 0.8759 $\pm$ \scriptsize{0.0463}
& 0.6670 $\pm$ \scriptsize{0.0940}
& 0.3438 $\pm$ \scriptsize{0.0534} \\

\proposed
\multirow{-7}{*}{\cellcolor{white}SSIM$\uparrow$}
& Self-supervised & Diffusion & SinoDiff
& \textbf{0.9730 $\pm$ \scriptsize{0.0040}}
& \textbf{0.9718 $\pm$ \scriptsize{0.0062}}
& \textbf{0.9715 $\pm$ \scriptsize{0.0057}}
& \textbf{0.9712 $\pm$ \scriptsize{0.0044}}
& \textbf{0.9679 $\pm$ \scriptsize{0.0054}}
& \textbf{0.9643 $\pm$ \scriptsize{0.0044}}
& \textbf{0.9561 $\pm$ \scriptsize{0.0093}}
& \textbf{0.9483 $\pm$ \scriptsize{0.0059}}
& \textbf{0.9275 $\pm$ \scriptsize{0.0050}} \\

\midrule

\oracle
\cellcolor{white}
& With Retraining & UNet & UNet
& 38.8046 $\pm$ \scriptsize{1.0661}
& 38.8557 $\pm$ \scriptsize{2.1839}
& 39.0800 $\pm$ \scriptsize{0.9907}
& 38.5874 $\pm$ \scriptsize{0.6322}
& 38.1414 $\pm$ \scriptsize{0.9193}
& 37.6762 $\pm$ \scriptsize{0.9439}
& 37.6710 $\pm$ \scriptsize{0.8954}
& 35.6225 $\pm$ \scriptsize{1.6570}
& 34.6569 $\pm$ \scriptsize{1.5389} \\

\oracle
\cellcolor{white}
& With Retraining & Diffusion & RED
& 40.8768 $\pm$ \scriptsize{0.5304}
& 37.9658 $\pm$ \scriptsize{4.3070}
& 38.9336 $\pm$ \scriptsize{0.9447}
& 35.5090 $\pm$ \scriptsize{6.7293}
& 37.2437 $\pm$ \scriptsize{2.8980}
& 35.2815 $\pm$ \scriptsize{2.2592}
& 37.2277 $\pm$ \scriptsize{1.9078}
& 28.7744 $\pm$ \scriptsize{6.5346}
& 24.3973 $\pm$ \scriptsize{5.0625} \\

\cmidrule(lr){2-13}

\wor
\cellcolor{white}
& W/O Retraining & UNet & UNet
& 34.7190 $\pm$ \scriptsize{2.3615}
& 34.4827 $\pm$ \scriptsize{1.9921}
& 34.6935 $\pm$ \scriptsize{2.1316}
& 34.5530 $\pm$ \scriptsize{2.3915}
& 34.7219 $\pm$ \scriptsize{2.2760}
& 34.4460 $\pm$ \scriptsize{1.9354}
& 34.3339 $\pm$ \scriptsize{1.8635}
& \underline{34.2713 $\pm$ \scriptsize{1.7318}}
& \underline{34.6569 $\pm$ \scriptsize{1.5389}} \\

\wor
\cellcolor{white}
& W/O Retraining & Diffusion & RED
& \underline{38.1020 $\pm$ \scriptsize{0.4024}}
& \underline{37.4127 $\pm$ \scriptsize{1.0501}}
& \underline{37.4584 $\pm$ \scriptsize{0.3592}}
& \underline{35.5616 $\pm$ \scriptsize{1.8487}}
& \underline{36.6754 $\pm$ \scriptsize{1.1260}}
& 33.4554 $\pm$ \scriptsize{3.1851}
& \underline{35.7257 $\pm$ \scriptsize{1.1954}}
& 29.3021 $\pm$ \scriptsize{5.2479}
& 24.3973 $\pm$ \scriptsize{5.0625} \\

\self
\cellcolor{white}
& Self-supervised & DIP & DIP
& 30.0118 $\pm$ \scriptsize{1.4024}
& 30.4390 $\pm$ \scriptsize{1.4434}
& 32.0342 $\pm$ \scriptsize{1.7580}
& 32.7332 $\pm$ \scriptsize{1.6303}
& 33.8781 $\pm$ \scriptsize{1.0414}
& \underline{35.0434 $\pm$ \scriptsize{2.1045}}
& 32.9420 $\pm$ \scriptsize{3.3338}
& 26.2971 $\pm$ \scriptsize{2.9244}
& 17.3277 $\pm$ \scriptsize{0.8931} \\

\self
\cellcolor{white}
& Self-supervised & DIP & ADMM-DIPTV
& 28.8484 $\pm$ \scriptsize{1.5592}
& 30.1704 $\pm$ \scriptsize{1.0337}
& 30.6030 $\pm$ \scriptsize{0.9227}
& 31.8338 $\pm$ \scriptsize{0.5415}
& 31.6526 $\pm$ \scriptsize{1.1923}
& 33.0166 $\pm$ \scriptsize{1.0206}
& 31.2558 $\pm$ \scriptsize{2.5104}
& 26.3818 $\pm$ \scriptsize{2.6870}
& 17.9793 $\pm$ \scriptsize{1.1202} \\

\proposed
\multirow{-7}{*}{\cellcolor{white}PSNR$\uparrow$}
& Self-supervised & Diffusion & SinoDiff
& \textbf{38.7322 $\pm$ \scriptsize{3.0172}}
& \textbf{38.7677 $\pm$ \scriptsize{2.6255}}
& \textbf{39.7028 $\pm$ \scriptsize{2.0534}}
& \textbf{41.1456 $\pm$ \scriptsize{0.4565}}
& \textbf{40.2970 $\pm$ \scriptsize{1.5303}}
& \textbf{39.8363 $\pm$ \scriptsize{1.2379}}
& \textbf{38.5963 $\pm$ \scriptsize{2.3486}}
& \textbf{37.9699 $\pm$ \scriptsize{0.9214}}
& \textbf{35.6861 $\pm$ \scriptsize{1.9407}} \\

\midrule

\oracle
\cellcolor{white}
& With Retraining & UNet & UNet
& 0.0069 $\pm$ \scriptsize{0.0011}
& 0.0067 $\pm$ \scriptsize{0.0024}
& 0.0068 $\pm$ \scriptsize{0.0016}
& 0.0074 $\pm$ \scriptsize{0.0009}
& 0.0082 $\pm$ \scriptsize{0.0028}
& 0.0094 $\pm$ \scriptsize{0.0018}
& 0.0091 $\pm$ \scriptsize{0.0014}
& 0.0154 $\pm$ \scriptsize{0.0055}
& 0.0190 $\pm$ \scriptsize{0.0042} \\

\oracle
\cellcolor{white}
& With Retraining & Diffusion & RED
& 0.0044 $\pm$ \scriptsize{0.0011}
& 0.0098 $\pm$ \scriptsize{0.0093}
& 0.0067 $\pm$ \scriptsize{0.0016}
& 0.0255 $\pm$ \scriptsize{0.0378}
& 0.0101 $\pm$ \scriptsize{0.0075}
& 0.0152 $\pm$ \scriptsize{0.0076}
& 0.0100 $\pm$ \scriptsize{0.0027}
& 0.0809 $\pm$ \scriptsize{0.0985}
& 0.1329 $\pm$ \scriptsize{0.0726} \\

\cmidrule(lr){2-13}

\wor
\cellcolor{white}
& W/O Retraining & UNet & UNet
& 0.0184 $\pm$ \scriptsize{0.0065}
& 0.0190 $\pm$ \scriptsize{0.0051}
& 0.0183 $\pm$ \scriptsize{0.0056}
& 0.0188 $\pm$ \scriptsize{0.0069}
& 0.0185 $\pm$ \scriptsize{0.0062}
& 0.0193 $\pm$ \scriptsize{0.0050}
& 0.0197 $\pm$ \scriptsize{0.0055}
& \underline{0.0202 $\pm$ \scriptsize{0.0049}}
& \underline{0.0190 $\pm$ \scriptsize{0.0042}} \\

\wor
\cellcolor{white}
& W/O Retraining & Diffusion & RED
& \underline{0.0075 $\pm$ \scriptsize{0.0005}}
& \underline{0.0095 $\pm$ \scriptsize{0.0022}}
& \underline{0.0089 $\pm$ \scriptsize{0.0020}}
& \underline{0.0157 $\pm$ \scriptsize{0.0095}}
& \underline{0.0110 $\pm$ \scriptsize{0.0030}}
& 0.0251 $\pm$ \scriptsize{0.0145}
& \underline{0.0141 $\pm$ \scriptsize{0.0035}}
& 0.0600 $\pm$ \scriptsize{0.0462}
& 0.1329 $\pm$ \scriptsize{0.0726} \\

\self
\cellcolor{white}
& Self-supervised & DIP & DIP
& 0.0257 $\pm$ \scriptsize{0.0082}
& 0.0239 $\pm$ \scriptsize{0.0046}
& 0.0189 $\pm$ \scriptsize{0.0047}
& 0.0182 $\pm$ \scriptsize{0.0063}
& 0.0157 $\pm$ \scriptsize{0.0044}
& \underline{0.0136 $\pm$ \scriptsize{0.0080}}
& 0.0190 $\pm$ \scriptsize{0.0157}
& 0.0675 $\pm$ \scriptsize{0.0484}
& 0.3180 $\pm$ \scriptsize{0.1144} \\

\self
\cellcolor{white}
& Self-supervised & DIP & ADMM-DIPTV
& 0.0357 $\pm$ \scriptsize{0.0143}
& 0.0296 $\pm$ \scriptsize{0.0100}
& 0.0275 $\pm$ \scriptsize{0.0080}
& 0.0235 $\pm$ \scriptsize{0.0062}
& 0.0261 $\pm$ \scriptsize{0.0115}
& 0.0221 $\pm$ \scriptsize{0.0108}
& 0.0277 $\pm$ \scriptsize{0.0208}
& 0.0684 $\pm$ \scriptsize{0.0484}
& 0.3012 $\pm$ \scriptsize{0.1217} \\

\proposed
\multirow{-7}{*}{\cellcolor{white}NMSE$\downarrow$}
& Self-supervised & Diffusion & SinoDiff
& \textbf{0.0064 $\pm$ \scriptsize{0.0023}}
& \textbf{0.0067 $\pm$ \scriptsize{0.0043}}
& \textbf{0.0057 $\pm$ \scriptsize{0.0030}}
& \textbf{0.0041 $\pm$ \scriptsize{0.0007}}
& \textbf{0.0052 $\pm$ \scriptsize{0.0016}}
& \textbf{0.0056 $\pm$ \scriptsize{0.0013}}
& \textbf{0.0080 $\pm$ \scriptsize{0.0045}}
& \textbf{0.0083 $\pm$ \scriptsize{0.0009}}
& \textbf{0.0138 $\pm$ \scriptsize{0.0046}} \\

\midrule

\oracle
\cellcolor{white}
& With Retraining & UNet & UNet
& 0.0015 $\pm$ \scriptsize{0.0002}
& 0.0014 $\pm$ \scriptsize{0.0002}
& 0.0017 $\pm$ \scriptsize{0.0003}
& 0.0020 $\pm$ \scriptsize{0.0003}
& 0.0022 $\pm$ \scriptsize{0.0002}
& 0.0027 $\pm$ \scriptsize{0.0003}
& 0.0031 $\pm$ \scriptsize{0.0004}
& 0.0039 $\pm$ \scriptsize{0.0005}
& 0.0048 $\pm$ \scriptsize{0.0007} \\

\oracle
\cellcolor{white}
& With Retraining & Diffusion & RED
& 0.0010 $\pm$ \scriptsize{0.0002}
& 0.0013 $\pm$ \scriptsize{0.0005}
& 0.0012 $\pm$ \scriptsize{0.0004}
& 0.0020 $\pm$ \scriptsize{0.0012}
& 0.0017 $\pm$ \scriptsize{0.0005}
& 0.0025 $\pm$ \scriptsize{0.0003}
& 0.0028 $\pm$ \scriptsize{0.0011}
& 0.0069 $\pm$ \scriptsize{0.0049}
& 0.0112 $\pm$ \scriptsize{0.0037} \\

\cmidrule(lr){2-13}

\wor
\cellcolor{white}
& W/O Retraining & UNet & UNet
& 0.0048 $\pm$ \scriptsize{0.0007}
& 0.0049 $\pm$ \scriptsize{0.0006}
& 0.0049 $\pm$ \scriptsize{0.0007}
& 0.0048 $\pm$ \scriptsize{0.0007}
& 0.0049 $\pm$ \scriptsize{0.0007}
& 0.0048 $\pm$ \scriptsize{0.0006}
& 0.0049 $\pm$ \scriptsize{0.0007}
& \underline{0.0049 $\pm$ \scriptsize{0.0006}}
& \underline{0.0048 $\pm$ \scriptsize{0.0007}} \\

\wor
\cellcolor{white}
& W/O Retraining & Diffusion & RED
& \underline{0.0018 $\pm$ \scriptsize{0.0004}}
& \underline{0.0020 $\pm$ \scriptsize{0.0005}}
& \underline{0.0020 $\pm$ \scriptsize{0.0004}}
& \underline{0.0024 $\pm$ \scriptsize{0.0003}}
& \underline{0.0024 $\pm$ \scriptsize{0.0006}}
& 0.0033 $\pm$ \scriptsize{0.0007}
& \underline{0.0033 $\pm$ \scriptsize{0.0008}}
& 0.0059 $\pm$ \scriptsize{0.0021}
& 0.0112 $\pm$ \scriptsize{0.0037} \\

\self
\cellcolor{white}
& Self-supervised & DIP & DIP
& 0.0033 $\pm$ \scriptsize{0.0012}
& 0.0032 $\pm$ \scriptsize{0.0012}
& 0.0024 $\pm$ \scriptsize{0.0006}
& 0.0030 $\pm$ \scriptsize{0.0010}
& 0.0032 $\pm$ \scriptsize{0.0012}
& \underline{0.0030 $\pm$ \scriptsize{0.0006}}
& 0.0085 $\pm$ \scriptsize{0.0043}
& 0.0240 $\pm$ \scriptsize{0.0062}
& 0.0558 $\pm$ \scriptsize{0.0055} \\

\self
\cellcolor{white}
& Self-supervised & DIP & ADMM-DIPTV
& 0.0065 $\pm$ \scriptsize{0.0013}
& 0.0064 $\pm$ \scriptsize{0.0006}
& 0.0064 $\pm$ \scriptsize{0.0011}
& 0.0067 $\pm$ \scriptsize{0.0009}
& 0.0069 $\pm$ \scriptsize{0.0009}
& 0.0081 $\pm$ \scriptsize{0.0011}
& 0.0138 $\pm$ \scriptsize{0.0046}
& 0.0315 $\pm$ \scriptsize{0.0073}
& 0.0598 $\pm$ \scriptsize{0.0051} \\

\proposed
\multirow{-7}{*}{\cellcolor{white}LPIPS$\downarrow$}
& Self-supervised & Diffusion & SinoDiff
& \textbf{0.0009 $\pm$ \scriptsize{0.0003}}
& \textbf{0.0009 $\pm$ \scriptsize{0.0001}}
& \textbf{0.0009 $\pm$ \scriptsize{0.0001}}
& \textbf{0.0009 $\pm$ \scriptsize{0.0002}}
& \textbf{0.0010 $\pm$ \scriptsize{0.0002}}
& \textbf{0.0012 $\pm$ \scriptsize{0.0003}}
& \textbf{0.0016 $\pm$ \scriptsize{0.0005}}
& \textbf{0.0020 $\pm$ \scriptsize{0.0003}}
& \textbf{0.0030 $\pm$ \scriptsize{0.0006}} \\

\bottomrule
\end{tabular}}
\end{table}

\begin{table}[t]
\centering
\caption{
Comparison of image-domain metrics between supervised methods with and without retraining,
self-supervised methods, and proposed SinoDiff across multiple dose levels for the
$^{18}$F-FDOPA dataset.
}
\label{tab:fdopa_all_dose}
\setlength{\tabcolsep}{2.5pt}

\resizebox{\textwidth}{!}{
\begin{tabular}{llllccccccccc}
\toprule
& & &
& \multicolumn{9}{c}{$^{18}$F-FDOPA Dose (\%)} \\
\cmidrule(lr){5-13}
Metric & Training setting & Backbone & Method
& 90 & 80 & 70 & 60 & 50 & 40 & 30 & 20 & 10 \\
\midrule

\oracle
\cellcolor{white}
& With Retraining & UNet & UNet
& 0.9239 $\pm$ \scriptsize{0.0094}
& 0.9211 $\pm$ \scriptsize{0.0128}
& 0.9146 $\pm$ \scriptsize{0.0115}
& 0.9091 $\pm$ \scriptsize{0.0131}
& 0.9035 $\pm$ \scriptsize{0.0118}
& 0.8971 $\pm$ \scriptsize{0.0193}
& 0.8776 $\pm$ \scriptsize{0.0211}
& 0.8577 $\pm$ \scriptsize{0.0273}
& 0.8360 $\pm$ \scriptsize{0.0230} \\

\oracle
\cellcolor{white}
& With Retraining & Diffusion & RED
& 0.9179 $\pm$ \scriptsize{0.0170}
& 0.9101 $\pm$ \scriptsize{0.0148}
& 0.8827 $\pm$ \scriptsize{0.0427}
& 0.8619 $\pm$ \scriptsize{0.0690}
& 0.8548 $\pm$ \scriptsize{0.0403}
& 0.8036 $\pm$ \scriptsize{0.1161}
& 0.7671 $\pm$ \scriptsize{0.1057}
& 0.7704 $\pm$ \scriptsize{0.0676}
& 0.6911 $\pm$ \scriptsize{0.0744} \\

\cmidrule(lr){2-13}

\wor
\cellcolor{white}
& W/O Retraining & UNet & UNet
& 0.8481 $\pm$ \scriptsize{0.0226}
& 0.8467 $\pm$ \scriptsize{0.0237}
& 0.8471 $\pm$ \scriptsize{0.0257}
& 0.8466 $\pm$ \scriptsize{0.0252}
& 0.8467 $\pm$ \scriptsize{0.0268}
& 0.8435 $\pm$ \scriptsize{0.0251}
& \underline{0.8475 $\pm$ \scriptsize{0.0187}}
& \textbf{0.8402 $\pm$ \scriptsize{0.0286}}
& \textbf{0.8360 $\pm$ \scriptsize{0.0230}} \\

\wor
\cellcolor{white}
& W/O Retraining & Diffusion & RED
& \underline{0.8994 $\pm$ \scriptsize{0.0080}}
& 0.8957 $\pm$ \scriptsize{0.0086}
& 0.8933 $\pm$ \scriptsize{0.0108}
& 0.8832 $\pm$ \scriptsize{0.0115}
& 0.8748 $\pm$ \scriptsize{0.0147}
& 0.8465 $\pm$ \scriptsize{0.0490}
& 0.8347 $\pm$ \scriptsize{0.0242}
& 0.7680 $\pm$ \scriptsize{0.0692}
& 0.6911 $\pm$ \scriptsize{0.0744} \\

\self
\cellcolor{white}
& Self-supervised & DIP & DIP
& 0.8903 $\pm$ \scriptsize{0.0469}
& \underline{0.9079 $\pm$ \scriptsize{0.0196}}
& \underline{0.9075 $\pm$ \scriptsize{0.0165}}
& \underline{0.8954 $\pm$ \scriptsize{0.0187}}
& \underline{0.8960 $\pm$ \scriptsize{0.0175}}
& \underline{0.8763 $\pm$ \scriptsize{0.0152}}
& 0.8457 $\pm$ \scriptsize{0.0179}
& 0.6408 $\pm$ \scriptsize{0.1444}
& 0.1893 $\pm$ \scriptsize{0.0855} \\

\self
\cellcolor{white}
& Self-supervised & DIP & ADMM-DIPTV
& 0.8330 $\pm$ \scriptsize{0.0405}
& 0.8276 $\pm$ \scriptsize{0.0175}
& 0.8530 $\pm$ \scriptsize{0.0168}
& 0.8183 $\pm$ \scriptsize{0.0491}
& 0.8239 $\pm$ \scriptsize{0.0451}
& 0.8161 $\pm$ \scriptsize{0.0485}
& 0.7934 $\pm$ \scriptsize{0.0440}
& 0.7733 $\pm$ \scriptsize{0.0303}
& 0.2730 $\pm$ \scriptsize{0.0594} \\

\proposed
\multirow{-7}{*}{\cellcolor{white}SSIM$\uparrow$}
& Self-supervised & Diffusion & SinoDiff
& \textbf{0.9155 $\pm$ \scriptsize{0.0335}}
& \textbf{0.9074 $\pm$ \scriptsize{0.0384}}
& \textbf{0.9112 $\pm$ \scriptsize{0.0254}}
& \textbf{0.8981 $\pm$ \scriptsize{0.0406}}
& \textbf{0.8812 $\pm$ \scriptsize{0.0475}}
& \textbf{0.8851 $\pm$ \scriptsize{0.0404}}
& \textbf{0.8762 $\pm$ \scriptsize{0.0224}}
& \underline{0.8391 $\pm$ \scriptsize{0.0475}}
& \underline{0.8259 $\pm$ \scriptsize{0.0112}} \\

\midrule

\oracle
\cellcolor{white}
& With Retraining & UNet & UNet
& 34.9929 $\pm$ \scriptsize{2.0972}
& 34.4290 $\pm$ \scriptsize{4.5906}
& 35.5686 $\pm$ \scriptsize{2.7479}
& 34.3992 $\pm$ \scriptsize{2.9532}
& 34.2544 $\pm$ \scriptsize{4.4771}
& 34.1367 $\pm$ \scriptsize{4.9132}
& 31.3616 $\pm$ \scriptsize{3.8304}
& 30.4065 $\pm$ \scriptsize{4.0996}
& 32.3569 $\pm$ \scriptsize{3.2097} \\

\oracle
\cellcolor{white}
& With Retraining & Diffusion & RED
& 37.1544 $\pm$ \scriptsize{3.0174}
& 36.4588 $\pm$ \scriptsize{2.3658}
& 33.6847 $\pm$ \scriptsize{6.1317}
& 32.4874 $\pm$ \scriptsize{6.1059}
& 31.3064 $\pm$ \scriptsize{4.3324}
& 29.7996 $\pm$ \scriptsize{6.0651}
& 29.0730 $\pm$ \scriptsize{6.6661}
& 29.3922 $\pm$ \scriptsize{4.3659}
& 27.4009 $\pm$ \scriptsize{3.4098} \\

\cmidrule(lr){2-13}

\wor
\cellcolor{white}
& W/O Retraining & UNet & UNet
& 31.4378 $\pm$ \scriptsize{4.2797}
& 31.1344 $\pm$ \scriptsize{4.4801}
& 31.1736 $\pm$ \scriptsize{4.1562}
& 31.1994 $\pm$ \scriptsize{4.1061}
& 31.6284 $\pm$ \scriptsize{4.6550}
& 30.7996 $\pm$ \scriptsize{4.2476}
& 32.1454 $\pm$ \scriptsize{3.3474}
& \underline{31.6373 $\pm$ \scriptsize{4.6615}}
& \underline{32.3569 $\pm$ \scriptsize{3.2097}} \\

\wor
\cellcolor{white}
& W/O Retraining & Diffusion & RED
& 34.7982 $\pm$ \scriptsize{2.4325}
& \textbf{35.1955 $\pm$ \scriptsize{1.3912}}
& \textbf{35.8308 $\pm$ \scriptsize{1.2473}}
& \textbf{34.6040 $\pm$ \scriptsize{1.3561}}
& \textbf{35.3845 $\pm$ \scriptsize{1.3242}}
& \underline{33.7730 $\pm$ \scriptsize{4.1059}}
& \underline{33.0336 $\pm$ \scriptsize{1.7884}}
& 29.4348 $\pm$ \scriptsize{4.2248}
& 27.4009 $\pm$ \scriptsize{3.4098} \\

\self
\cellcolor{white}
& Self-supervised & DIP & DIP
& 29.4958 $\pm$ \scriptsize{4.3894}
& 31.3439 $\pm$ \scriptsize{1.4700}
& 31.7756 $\pm$ \scriptsize{1.5416}
& 30.8086 $\pm$ \scriptsize{1.1403}
& 33.0639 $\pm$ \scriptsize{1.8954}
& 30.3725 $\pm$ \scriptsize{1.0298}
& 30.2080 $\pm$ \scriptsize{1.2191}
& 23.7929 $\pm$ \scriptsize{4.7442}
& 16.9082 $\pm$ \scriptsize{1.0582} \\

\self
\cellcolor{white}
& Self-supervised & DIP & ADMM-DIPTV
& 27.4759 $\pm$ \scriptsize{3.2066}
& 25.5696 $\pm$ \scriptsize{1.9780}
& 27.4050 $\pm$ \scriptsize{2.7005}
& 24.8905 $\pm$ \scriptsize{2.9014}
& 26.5052 $\pm$ \scriptsize{3.7304}
& 25.1383 $\pm$ \scriptsize{2.4355}
& 25.9057 $\pm$ \scriptsize{2.6334}
& 29.3154 $\pm$ \scriptsize{2.1662}
& 19.4562 $\pm$ \scriptsize{1.4518} \\

\proposed
\multirow{-7}{*}{\cellcolor{white}PSNR$\uparrow$}
& Self-supervised & Diffusion & SinoDiff
& \textbf{35.6973 $\pm$ \scriptsize{5.7850}}
& \underline{34.8176 $\pm$ \scriptsize{6.8346}}
& \underline{34.4763 $\pm$ \scriptsize{4.4405}}
& \underline{33.9884 $\pm$ \scriptsize{5.2485}}
& \underline{33.1930 $\pm$ \scriptsize{6.1360}}
& \textbf{34.4003 $\pm$ \scriptsize{5.3196}}
& \textbf{33.6801 $\pm$ \scriptsize{2.3321}}
& \textbf{31.4825 $\pm$ \scriptsize{4.8337}}
& \textbf{32.5605 $\pm$ \scriptsize{3.9046}} \\

\midrule

\oracle
\cellcolor{white}
& With Retraining & UNet & UNet
& 0.0250 $\pm$ \scriptsize{0.0066}
& 0.0295 $\pm$ \scriptsize{0.0145}
& 0.0245 $\pm$ \scriptsize{0.0085}
& 0.0289 $\pm$ \scriptsize{0.0106}
& 0.0314 $\pm$ \scriptsize{0.0156}
& 0.0346 $\pm$ \scriptsize{0.0234}
& 0.0509 $\pm$ \scriptsize{0.0254}
& 0.0634 $\pm$ \scriptsize{0.0378}
& 0.0513 $\pm$ \scriptsize{0.0247} \\

\oracle
\cellcolor{white}
& With Retraining & Diffusion & RED
& 0.0206 $\pm$ \scriptsize{0.0115}
& 0.0235 $\pm$ \scriptsize{0.0138}
& 0.0454 $\pm$ \scriptsize{0.0385}
& 0.0589 $\pm$ \scriptsize{0.0593}
& 0.0590 $\pm$ \scriptsize{0.0350}
& 0.0949 $\pm$ \scriptsize{0.0986}
& 0.1150 $\pm$ \scriptsize{0.0989}
& 0.0942 $\pm$ \scriptsize{0.0614}
& 0.1350 $\pm$ \scriptsize{0.0866} \\

\cmidrule(lr){2-13}

\wor
\cellcolor{white}
& W/O Retraining & UNet & UNet
& 0.0611 $\pm$ \scriptsize{0.0379}
& 0.0648 $\pm$ \scriptsize{0.0417}
& 0.0637 $\pm$ \scriptsize{0.0437}
& 0.0633 $\pm$ \scriptsize{0.0419}
& 0.0614 $\pm$ \scriptsize{0.0455}
& 0.0676 $\pm$ \scriptsize{0.0419}
& 0.0521 $\pm$ \scriptsize{0.0256}
& \underline{0.0621 $\pm$ \scriptsize{0.0458}}
& \textbf{0.0513 $\pm$ \scriptsize{0.0247}} \\

\wor
\cellcolor{white}
& W/O Retraining & Diffusion & RED
& \textbf{0.0284 $\pm$ \scriptsize{0.0093}}
& \textbf{0.0268 $\pm$ \scriptsize{0.0027}}
& \textbf{0.0245 $\pm$ \scriptsize{0.0055}}
& \textbf{0.0314 $\pm$ \scriptsize{0.0100}}
& \textbf{0.0293 $\pm$ \scriptsize{0.0033}}
& \underline{0.0462 $\pm$ \scriptsize{0.0363}}
& \underline{0.0474 $\pm$ \scriptsize{0.0128}}
& 0.0959 $\pm$ \scriptsize{0.0607}
& 0.1350 $\pm$ \scriptsize{0.0866} \\

\self
\cellcolor{white}
& Self-supervised & DIP & DIP
& 0.0724 $\pm$ \scriptsize{0.0637}
& 0.0426 $\pm$ \scriptsize{0.0086}
& 0.0408 $\pm$ \scriptsize{0.0107}
& 0.0496 $\pm$ \scriptsize{0.0159}
& \underline{0.0366 $\pm$ \scriptsize{0.0133}}
& 0.0589 $\pm$ \scriptsize{0.0120}
& 0.0689 $\pm$ \scriptsize{0.0156}
& 0.2150 $\pm$ \scriptsize{0.1397}
& 0.6981 $\pm$ \scriptsize{0.1267} \\
\self
\cellcolor{white}
& Self-supervised & DIP & ADMM-DIPTV
& 0.0986 $\pm$ \scriptsize{0.0502}
& 0.1236 $\pm$ \scriptsize{0.0350}
& 0.0910 $\pm$ \scriptsize{0.0308}
& 0.1473 $\pm$ \scriptsize{0.0655}
& 0.1277 $\pm$ \scriptsize{0.0769}
& 0.1570 $\pm$ \scriptsize{0.0742}
& 0.1666 $\pm$ \scriptsize{0.0797}
& 0.1017 $\pm$ \scriptsize{0.0486}
& 0.6037 $\pm$ \scriptsize{0.1288} \\

\proposed
\multirow{-7}{*}{\cellcolor{white}NMSE$\downarrow$}
& Self-supervised & Diffusion & SinoDiff
& \underline{0.0315 $\pm$ \scriptsize{0.0346}}
& \underline{0.0365 $\pm$ \scriptsize{0.0354}}
& \underline{0.0312 $\pm$ \scriptsize{0.0209}}
& \underline{0.0383 $\pm$ \scriptsize{0.0347}}
& 0.0489 $\pm$ \scriptsize{0.0456}
& \textbf{0.0383 $\pm$ \scriptsize{0.0327}}
& \textbf{0.0376 $\pm$ \scriptsize{0.0160}}
& \textbf{0.0607 $\pm$ \scriptsize{0.0382}}
& \underline{0.0516 $\pm$ \scriptsize{0.0259}} \\

\midrule

\oracle
\cellcolor{white}
& With Retraining & UNet & UNet
& 0.0021 $\pm$ \scriptsize{0.0007}
& 0.0021 $\pm$ \scriptsize{0.0003}
& 0.0023 $\pm$ \scriptsize{0.0003}
& 0.0025 $\pm$ \scriptsize{0.0003}
& 0.0029 $\pm$ \scriptsize{0.0004}
& 0.0034 $\pm$ \scriptsize{0.0004}
& 0.0041 $\pm$ \scriptsize{0.0004}
& 0.0051 $\pm$ \scriptsize{0.0005}
& 0.0062 $\pm$ \scriptsize{0.0006} \\

\oracle
\cellcolor{white}
& With Retraining & Diffusion & RED
& 0.0022 $\pm$ \scriptsize{0.0004}
& 0.0025 $\pm$ \scriptsize{0.0004}
& 0.0035 $\pm$ \scriptsize{0.0013}
& 0.0044 $\pm$ \scriptsize{0.0023}
& 0.0047 $\pm$ \scriptsize{0.0013}
& 0.0070 $\pm$ \scriptsize{0.0044}
& 0.0090 $\pm$ \scriptsize{0.0046}
& 0.0087 $\pm$ \scriptsize{0.0024}
& 0.0132 $\pm$ \scriptsize{0.0038} \\

\cmidrule(lr){2-13}

\wor
\cellcolor{white}
& W/O Retraining & UNet & UNet
& 0.0076 $\pm$ \scriptsize{0.0006}
& 0.0076 $\pm$ \scriptsize{0.0006}
& 0.0076 $\pm$ \scriptsize{0.0006}
& 0.0075 $\pm$ \scriptsize{0.0005}
& 0.0074 $\pm$ \scriptsize{0.0005}
& 0.0074 $\pm$ \scriptsize{0.0006}
& 0.0071 $\pm$ \scriptsize{0.0005}
& 0.0070 $\pm$ \scriptsize{0.0005}
& \underline{0.0062 $\pm$ \scriptsize{0.0006}} \\

\wor
\cellcolor{white}
& W/O Retraining & Diffusion & RED
& \underline{0.0032 $\pm$ \scriptsize{0.0004}}
& \underline{0.0032 $\pm$ \scriptsize{0.0006}}
& \underline{0.0033 $\pm$ \scriptsize{0.0005}}
& \underline{0.0035 $\pm$ \scriptsize{0.0005}}
& \underline{0.0040 $\pm$ \scriptsize{0.0007}}
& \underline{0.0052 $\pm$ \scriptsize{0.0021}}
& \underline{0.0057 $\pm$ \scriptsize{0.0008}}
& \underline{0.0090 $\pm$ \scriptsize{0.0027}}
& 0.0132 $\pm$ \scriptsize{0.0038} \\

\self
\cellcolor{white}
& Self-supervised & DIP & DIP
& 0.0062 $\pm$ \scriptsize{0.0035}
& 0.0045 $\pm$ \scriptsize{0.0020}
& 0.0041 $\pm$ \scriptsize{0.0023}
& 0.0051 $\pm$ \scriptsize{0.0029}
& 0.0048 $\pm$ \scriptsize{0.0023}
& 0.0061 $\pm$ \scriptsize{0.0027}
& 0.0063 $\pm$ \scriptsize{0.0027}
& 0.0165 $\pm$ \scriptsize{0.0048}
& 0.0789 $\pm$ \scriptsize{0.0107} \\

\self
\cellcolor{white}
& Self-supervised & DIP & ADMM-DIPTV
& 0.0112 $\pm$ \scriptsize{0.0018}
& 0.0111 $\pm$ \scriptsize{0.0006}
& 0.0101 $\pm$ \scriptsize{0.0006}
& 0.0111 $\pm$ \scriptsize{0.0017}
& 0.0115 $\pm$ \scriptsize{0.0015}
& 0.0115 $\pm$ \scriptsize{0.0016}
& 0.0143 $\pm$ \scriptsize{0.0015}
& 0.0210 $\pm$ \scriptsize{0.0045}
& 0.0782 $\pm$ \scriptsize{0.0063} \\

\proposed
\multirow{-7}{*}{\cellcolor{white}LPIPS$\downarrow$}
& Self-supervised & Diffusion & SinoDiff
& \textbf{0.0026 $\pm$ \scriptsize{0.0011}}
& \textbf{0.0027 $\pm$ \scriptsize{0.0010}}
& \textbf{0.0024 $\pm$ \scriptsize{0.0006}}
& \textbf{0.0027 $\pm$ \scriptsize{0.0011}}
& \textbf{0.0033 $\pm$ \scriptsize{0.0018}}
& \textbf{0.0029 $\pm$ \scriptsize{0.0012}}
& \textbf{0.0029 $\pm$ \scriptsize{0.0006}}
& \textbf{0.0042 $\pm$ \scriptsize{0.0017}}
& \textbf{0.0038 $\pm$ \scriptsize{0.0008}} \\

\bottomrule
\end{tabular}}
\end{table}

\begin{figure}
\centering
\includegraphics[width=0.49\linewidth]{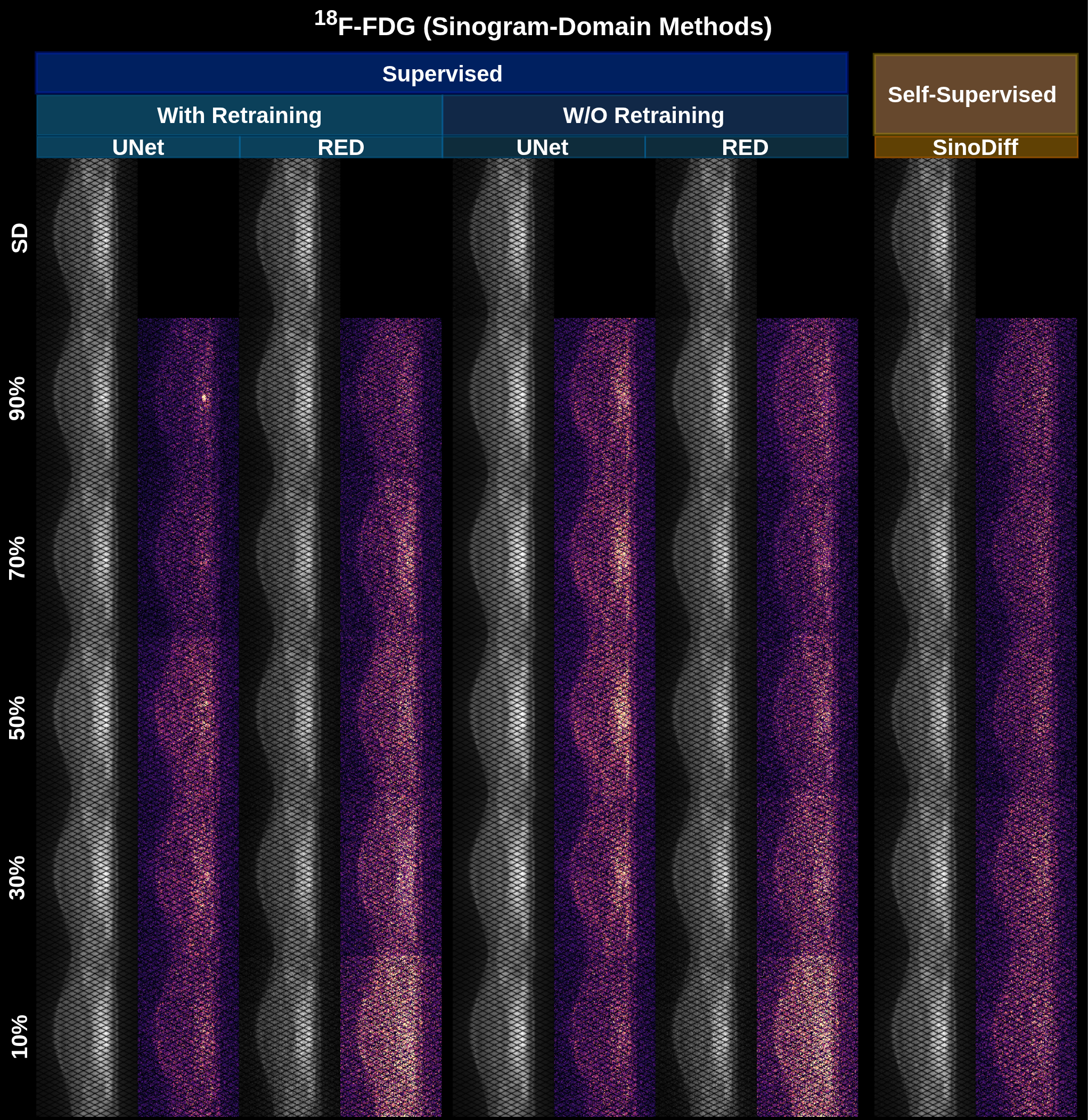} 
\includegraphics[width=0.49\linewidth]{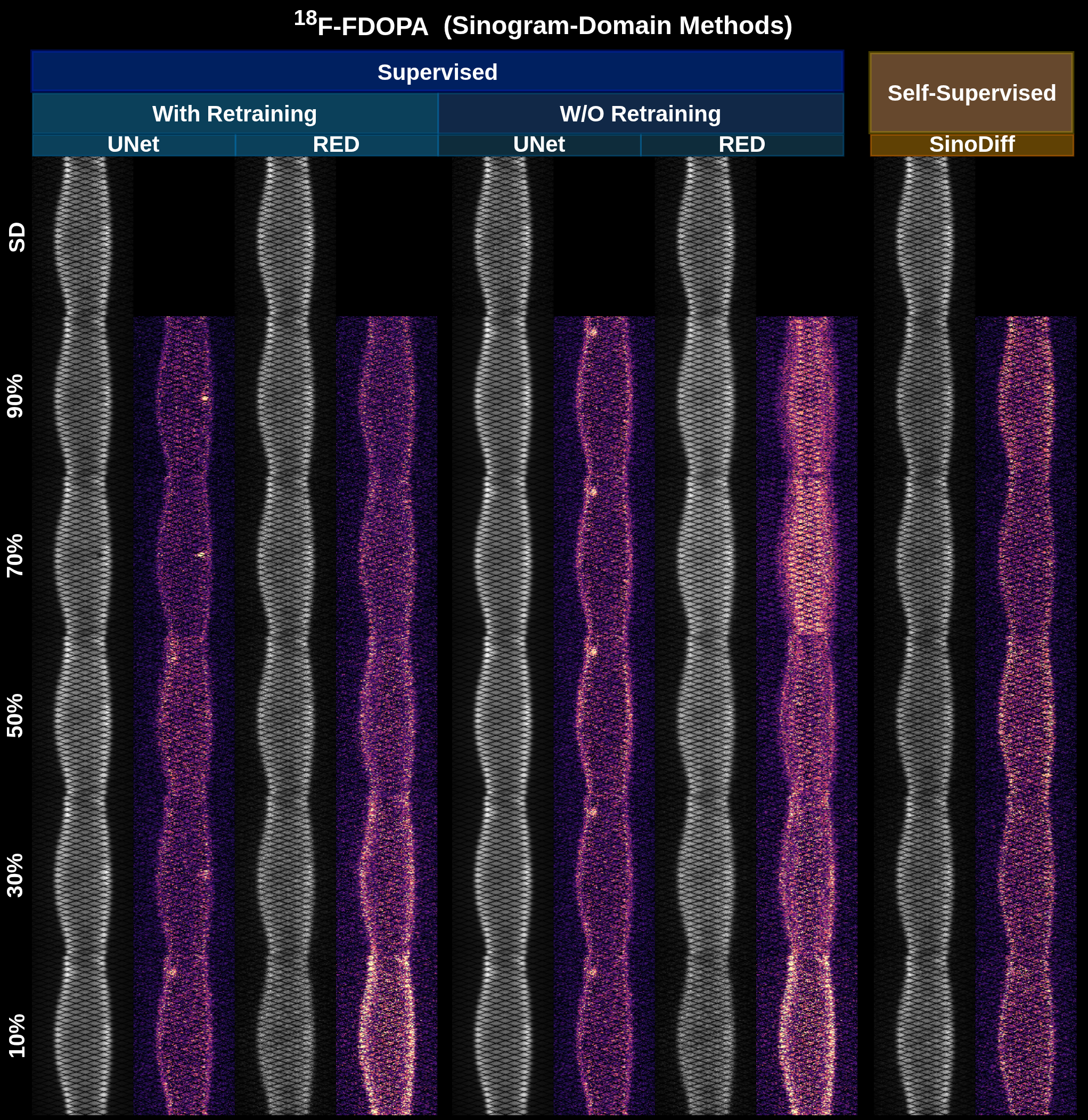}
\caption{Sinogram-domain qualitative comparison of supervised sinogram-based methods with and without retraining, and proposed SinoDiff on $^{18}$F-FDG and $^{18}$F-FDOPA datasets across multiple dose levels. Grayscale images show sinograms (display range: 0.0–1.0), while colour maps show absolute errors ($^{18}$F-FDG range: 0.0–0.11; $^{18}$F-FDOPA range: 0.0–0.15).}
\label{fig:dose_comp_sino}
\end{figure}

\paragraph{Sinogram Domain Analysis:} 

Fig.~\ref{fig:dose_comp_sino} presents qualitative sinogram-domain comparisons and corresponding absolute error maps across multiple dose levels for the $^{18}$F-FDG and $^{18}$F-FDOPA datasets. Supervised methods retrained independently for each dose level achieve strong quality, particularly at higher dose levels. However, when evaluated without retraining, both UNet and RED exhibit noticeable degradation and increased residual artefacts, especially in lower-dose regimes. In contrast, SinoDiff maintains stable quality across all predefined dose levels using a single unified model. The improvement becomes more apparent in the error maps, where baseline methods show amplified local residual artefacts and structural inconsistencies, particularly for the noisier $^{18}$F-FDOPA dataset.

Table~\ref{tab:fdg_fdopa_all_dose_sino} summarises quantitative sinogram-domain count consistency metrics, including mean absolute ACB, mean absolute RCB, and mean GCB. Across both datasets, SinoDiff consistently achieves the lowest bias values across most dose levels, indicating improved preservation of global count statistics and relative count consistency. The performance gap becomes more pronounced at lower dose levels, where competing methods exhibit rapidly increasing count bias and structural inconsistency. Although at the 10\% dose level, supervised baseline with and without retraining achieve competitive or second-best performance, these models were explicitly trained for that specific dose level. In contrast, SinoDiff maintains stable performance across all predefined dose levels using a single unified model without retraining. These results suggest that incorporating PET acquisition statistics directly within the diffusion process improves not only image-domain quality but also sinogram-domain consistency.

\begin{table}[t]
\centering
\caption{
Sinogram-domain metric comparison of supervised sinogram-based methods with and without retraining, and the proposed SinoDiff, across multiple dose levels on the $^{18}$F-FDG and $^{18}$F-FDOPA datasets.
}
\label{tab:fdg_fdopa_all_dose_sino}
\setlength{\tabcolsep}{2.5pt}

\resizebox{\textwidth}{!}{
\begin{tabular}{llllccccccccc ccccccccc}
\toprule
& & &
& \multicolumn{9}{c}{$^{18}$F-FDG Dose (\%)}
& \multicolumn{9}{c}{$^{18}$F-FDOPA Dose (\%)} \\
\cmidrule(lr){5-13}
\cmidrule(lr){14-22}

Metric & Training setting & Backbone & Method
& 90 & 80 & 70 & 60 & 50 & 40 & 30 & 20 & 10
& 90 & 80 & 70 & 60 & 50 & 40 & 30 & 20 & 10 \\
\midrule

\oracle
\cellcolor{white}
& With Retraining & UNet & UNet
& 1.7320 & 0.7053 & 1.0833 & 0.8582 & 0.6935 & 0.6747 & 0.6765 & 1.6874 & 1.3881
& 0.9863 & 1.1621 & 0.6021 & 3.5137 & 3.6351 & 4.1192 & 1.1374 & 1.8169 & 5.5261 \\

\oracle
\cellcolor{white}
& With Retraining & Diffusion & RED
& 0.7581 & 0.4777 & 5.0901 & 1.3727 & 0.6774 & 0.5226 & 0.9061 & 1.0461 & 3.7352
& 3.5738 & 4.6131 & 0.7527 & 1.0155 & 1.7048 & 3.1004 & 7.3461 & 9.9194 & 9.2048 \\

\cmidrule(lr){2-22}

\wor
\cellcolor{white}ACB$\downarrow$
& W/O Retraining & UNet & UNet
& \underline{5.9137} & \underline{5.8288} & \underline{5.7092} & \underline{5.5530}
& \underline{5.3205} & \underline{5.0455} & \underline{4.5773} & 3.6605
& \textbf{1.3881}
& \underline{10.4488} & \underline{10.4000} & \underline{10.2349}
& \underline{10.1205} & \underline{9.8753} & \underline{9.5760}
& \underline{9.0334} & \underline{8.0658} & \underline{5.5261} \\

\wor
\cellcolor{white}
& W/O Retraining & Diffusion & RED
& 13.8571 & 13.0920 & 12.5430 & 11.4675 & 10.3924 & 8.8767
& 6.4185 & \underline{3.0231} & 3.7917
& 41.0297 & 40.0125 & 38.6027 & 36.8995 & 35.0082 & 31.7149
& 27.8344 & 21.5730 & 9.2048 \\

\proposed
\cellcolor{white}
& Self-supervised & Diffusion & SinoDiff
& \textbf{0.0394} & \textbf{0.0693} & \textbf{0.1127}
& \textbf{0.1685} & \textbf{0.2473} & \textbf{0.3666}
& \textbf{0.5518} & \textbf{0.8810} & \underline{1.5438}
& \textbf{0.0978} & \textbf{0.2132} & \textbf{0.3621}
& \textbf{0.5442} & \textbf{0.7471} & \textbf{0.9893}
& \textbf{1.3121} & \textbf{1.8217} & \textbf{2.9275} \\

\midrule

\oracle
\cellcolor{white}
& With Retraining & UNet & UNet
& 1.5330 & 0.8278 & 0.9175 & 1.5012 & 0.8415 & 1.4351 & 0.9364 & 2.5534 & 2.5468
& 2.4191 & 2.2641 & 1.7814 & 3.3439 & 2.7564 & 3.8955 & 2.8043 & 2.7824 & 6.7473 \\

\oracle
\cellcolor{white}
& With Retraining & Diffusion & RED
& 1.5242 & 2.0126 & 12.3814 & 3.2274 & 0.9547 & 2.5083 & 4.7625 & 5.5182 & 3.9256
& 6.9957 & 11.7577 & 1.7257 & 1.2580 & 2.4734 & 7.5994 & 12.0793 & 22.7070 & 14.7529 \\

\cmidrule(lr){2-22}

\wor
\cellcolor{white}RCB$\downarrow$
& W/O Retraining & UNet & UNet
& \underline{7.5139} & \underline{7.4219} & \underline{7.3046} & \underline{7.1268}
& \underline{6.8799} & \underline{6.5796} & \underline{6.0728} & 5.0678
& \underline{2.5468}
& \underline{12.3972} & \underline{12.3406} & \underline{12.1759}
& \underline{12.0403} & \underline{11.7744} & \underline{11.4512}
& \underline{10.8401} & \underline{9.7438} & \underline{6.7473} \\

\wor
\cellcolor{white}
& W/O Retraining & Diffusion & RED
& 13.2682 & 12.5580 & 12.0111 & 11.0209 & 10.0369 & 8.6209
& 6.2525 & \underline{3.0348} & 3.9853
& 44.0940 & 43.0994 & 41.8587 & 40.3593 & 38.6810 & 35.7263
& 32.2986 & 26.5654 & 14.7529 \\

\proposed
\cellcolor{white}
& Self-supervised & Diffusion & SinoDiff
& \textbf{0.0782} & \textbf{0.1508} & \textbf{0.2474}
& \textbf{0.3621} & \textbf{0.5039} & \textbf{0.6986}
& \textbf{0.9634} & \textbf{1.4161} & \textbf{2.3069}
& \textbf{0.1699} & \textbf{0.3519} & \textbf{0.5773}
& \textbf{0.8586} & \textbf{1.1858} & \textbf{1.6073}
& \textbf{2.1543} & \textbf{2.9514} & \textbf{4.2882} \\

\midrule

\oracle
\cellcolor{white}
& With Retraining & UNet & UNet
& 1.7325 & 0.6195 & 1.0819 & 0.8126 & -0.5723 & 0.3268 & -0.5642 & 1.6958 & 1.3781
& 0.9494 & 1.1572 & -0.4519 & 3.5175 & 3.6376 & 4.1256 & 1.1211 & 1.8113 & 5.5298 \\

\oracle
\cellcolor{white}
& With Retraining & Diffusion & RED
& 0.7370 & -0.1865 & 5.0524 & 1.3149 & -0.5010 & -0.2393 & -0.3919 & 0.8431 & -3.7193
& 3.5510 & 4.5702 & -0.5120 & 0.8932 & 1.6918 & 2.9884 & 7.3118 & 9.8558 & 9.1691 \\

\cmidrule(lr){2-22}

\wor
\cellcolor{white}GCB$\rightarrow 0$
& W/O Retraining & UNet & UNet
& \underline{5.9187} & \underline{5.8336} & \underline{5.7142}
& \underline{5.5577} & \underline{5.3255} & \underline{5.0503}
& \underline{4.5821} & 3.6655 & \textbf{1.3781}
& \underline{10.4507} & \underline{10.4018} & \underline{10.2368}
& \underline{10.1225} & \underline{9.8771} & \underline{9.5777}
& \underline{9.0350} & \underline{8.0685} & \underline{5.5298} \\

\wor
\cellcolor{white}
& W/O Retraining & Diffusion & RED
& 13.8950 & 13.1300 & 12.5802 & 11.5043 & 10.4277 & 8.9107
& 6.4437 & \underline{3.0167} & -3.7760
& 40.9902 & 39.9738 & 38.5632 & 36.8603 & 34.9687 & 31.6762
& 27.7945 & 21.5345 & 9.1691 \\

\proposed
\cellcolor{white}
& Self-supervised & Diffusion & SinoDiff
& \textbf{0.0321} & \textbf{0.0664} & \textbf{0.1123}
& \textbf{0.1680} & \textbf{0.2465} & \textbf{0.3650}
& \textbf{0.5494} & \textbf{0.8770} & \underline{1.5388}
& \textbf{0.0938} & \textbf{0.2084} & \textbf{0.3546}
& \textbf{0.5336} & \textbf{0.7345} & \textbf{0.9736}
& \textbf{1.2929} & \textbf{1.7978} & \textbf{2.9023} \\

\bottomrule
\end{tabular}
}

\vspace{1mm}


\end{table}

\begin{table}[t]
\centering
\scriptsize
\caption{Ablation study for SinoDiff under different architectural configurations using 10-step inference at 10\% dose on the $^{18}$F-FDG dataset. SSIM, PSNR, NMSE, and LPIPS assess image quality, while ACB, RCB, and GCB evaluate sinogram domain count consistency.}
\label{tab:ablation_sinodiff}
\resizebox{\textwidth}{!}{
\begin{tabular}{c c c c | c c c c | c c c}
\hline
\multirow{2}{*}{Version} 
& \multirow{2}{*}{CM} 
& \multirow{2}{*}{FDC} 
& \multirow{2}{*}{$\mathcal{L}_{Freq}$} 
& \multicolumn{4}{c|}{Image-domain Metrics} 
& \multicolumn{3}{c}{Sinogram Domain Count Consistency} \\
\cline{5-8}
\cline{9-11}
& & & 
& SSIM $\uparrow$ 
& PSNR $\uparrow$ 
& NMSE $\downarrow$ 
& LPIPS $\downarrow$
& Mean $|\mathrm{ACB}|$ $\downarrow$ 
& Mean $|\mathrm{RCB}|$ $\downarrow$ 
& GCB $\rightarrow 0$ \\
\hline

V1   
& \xmark  
& \xmark  
& \xmark  
& 0.9161$\pm$0.0070 
& 35.0609$\pm$0.9545 
& 0.0140$\pm$0.0019 
& 0.0039$\pm$0.0006
& 2.0809$\pm$0.0859 
& 4.6096$\pm$0.0628 
& 2.0399$\pm$0.0850 \\

V2   
& \cmark 
& \xmark  
& \xmark  
& 0.9240$\pm$0.0059 
& 35.1621$\pm$1.5892 
& 0.0139$\pm$0.0033 
& 0.0032$\pm$0.0005
& 2.6693$\pm$0.0732 
& 4.8414$\pm$0.0532 
& 2.6437$\pm$0.0696 \\

V3   
& \cmark 
& \cmark 
& \xmark  
& \underline{0.9259$\pm$0.0084} 
& \textbf{35.8390$\pm$0.7006} 
& \textbf{0.0130$\pm$0.0029} 
& \underline{0.0032$\pm$0.0004}
& \underline{1.7728$\pm$0.0722} 
& \underline{2.6512$\pm$0.1858} 
& \underline{1.7620$\pm$0.0716} \\

SinoDiff   
& \cmark 
& \cmark 
& \cmark  
& \textbf{0.9275$\pm$0.0050} 
& \underline{35.6861$\pm$1.9407} 
& \underline{0.0138$\pm$0.0046} 
& \textbf{0.0030$\pm$0.0006}
& \textbf{1.5438$\pm$0.0561} 
& \textbf{2.3069$\pm$0.0670} 
& \textbf{1.5388$\pm$0.0532} \\

\hline
\end{tabular}
}
\end{table}

\paragraph{\textbf{Ablation Study.}}
To evaluate the contribution of individual components, we conducted an ablation study with multiple SinoDiff variants: 
\textbf{V1:} SinoDiff model using simple UNet; \textbf{V2:} SinoDiff model using UNet with channel mixing; \textbf{V3:} SinoDiff model using UNet with channel mixing and FDC module; \textbf{Proposed:} Full SinoDiff model.
Quantitative results at the 10\% dose level on $^{18}$F-FDG dataset are summarised in Table~\ref{tab:ablation_sinodiff}. Introducing channel mixing improves inter-slice consistency and image quality, while incorporating the FDC module further improves both image-domain quality and sinogram-domain count consistency. The full SinoDiff model achieves the best overall balance across image-domain metrics and sinogram-domain consistency measures, demonstrating the complementary contribution of frequency-aware modelling.

\paragraph{Generalisation to unseen intermediate dose levels.}
To evaluate whether SinoDiff was limited to the predefined training dose steps, we tested the trained model on unseen intermediate dose levels that were not used during training. Specifically, we evaluated performance at 15\%, 45\%, and 75\% dose levels, representing low, mid-range, and high intermediate-dose conditions. As shown in Table~\ref{tab:unseen_dose_results}, SinoDiff maintained stable performance across these unseen dose levels for both FDG and FDOPA datasets. These results indicate that exact matching to the predefined training dose steps is not required at inference, and that the proposed progressive dose-recovery framework can generalise to intermediate dose levels.

\begin{table}[t]
\centering
\tiny
\caption{SinoDiff performance on unseen/intermediate dose levels for $^{18}$F-FDG dataset.}
\label{tab:unseen_dose_results}
\resizebox{\linewidth}{!}{
\begin{tabular}{lcccccccc}
\hline
\multirow{2}{*}{Dose} 
& \multicolumn{4}{c}{FDG} 
& \multicolumn{4}{c}{FDOPA} \\
\cmidrule(lr){2-5} \cmidrule(lr){6-9}
& SSIM$\uparrow$ & PSNR$\uparrow$ & NMSE$\downarrow$ & LPIPS$\downarrow$
& SSIM$\uparrow$ & PSNR$\uparrow$ & NMSE$\downarrow$ & LPIPS$\downarrow$ \\
\midrule
15\% & 0.9283 $\pm$ 0.0046
     & 35.13 $\pm$ 1.18
     & 0.0137 $\pm$ 0.0036
     & 0.0038 $\pm$ 0.0007 
     & 0.8604 $\pm$ 0.0285
     & 33.13 $\pm$ 2.98
     & 0.0430 $\pm$ 0.0193
     & 0.0034 $\pm$ 0.0011\\
45\% & 0.9660 $\pm$ 0.0047
     & 39.57 $\pm$ 1.08
     & 0.0054 $\pm$ 0.0011
     & 0.0012 $\pm$ 0.0003 
     & 0.8929 $\pm$ 0.0336
     & 35.38 $\pm$ 4.61
     & 0.0315 $\pm$ 0.0254
     & 0.0025 $\pm$ 0.0010 \\
75\% & 0.9739 $\pm$ 0.0049
     & 41.31 $\pm$ 1.32
     & 0.0040 $\pm$ 0.0020
     & 0.0008 $\pm$ 0.0001 
     & 0.9126 $\pm$ 0.0326
     & 35.52 $\pm$ 6.27
     & 0.0313 $\pm$ 0.0297
     & 0.0024 $\pm$ 0.0008 \\
\hline
\end{tabular}
}
\end{table}

\paragraph{\textbf{Discussion and Limitations.}}
The results show that incorporating PET acquisition physics into a self-supervised diffusion framework enables robust LD-to-SD recovery. By using Poisson thinning and progressive count restoration, SinoDiff maintains dose-consistent performance across a wide range of dose levels. It achieves this without retraining, demonstrating competitive performance with supervised methods and outperforming self-supervised methods.

The improved signal recovery of SinoDiff stems from its incremental count recovery grounded in the underlying PET signal acquisition process. 
In contrast, although UNet and RED are fully supervised, their learning strategies are fundamentally different from SinoDiff. UNet performs one-shot LD to SD mapping, which becomes unstable at very low doses. RED introduces diffusion, but still predicts a large residual toward the final SD image at each step, which remains a complex and unstable task.
Self-supervised methods are highly sensitive to optimisation and rely on carefully tuned stopping~\cite{nittscher2024svd, shi2022measuring}, which makes them impractical to achieve in simple settings and results in inconsistent performance across dose levels.

One major limitation for SinoDiff is that it operates on predefined dose steps and still requires SD sinograms, which may limit flexibility in true LD-only clinical scenarios. Future work will address finer dose scheduling and reduced SD dependence. As the datasets do not provide pathology-specific information or clinical reader evaluation, the reported results represent technical image-quality assessment rather than clinical validation; pathology-based and reader studies remain important future work.

\section{Conclusion}
\label{sec:conclusion}

We introduced SinoDiff, a physics-consistent self-supervised diffusion framework for unified LD to SD PET sinogram recovery. By embedding the PET data model into the forward diffusion process, SinoDiff aligns generative modelling with the underlying PET acquisition statistics, which enables a progressive and physically grounded count recovery process. The proposed FDC further enhances global sinogram consistency by modelling long-range projection dependencies.

A single trained SinoDiff model generalises across multiple predefined dose levels without paired LD–SD supervision or retraining. Experiments on two clinical PET datasets demonstrate consistent improvements over both supervised and self-supervised baselines in both the image and sinogram domains, particularly in very low dose settings.
Future work will focus on adaptive dose scheduling for broader clinical applicability.

\bibliography{references}
\end{document}